\documentclass[acmtog]{acmart}
\acmSubmissionID{1784}

\usepackage{color}
\definecolor{olive}{rgb}{0.1,0.8,0.3}
\definecolor{mauve}{rgb}{0.48,0,0.72}

\usepackage{multirow}
\usepackage{bm}

\usepackage{mathtools}

\DeclarePairedDelimiter\floor{\lfloor}{\rfloor}

\usepackage[ruled]{algorithm2e} 

\SetAlFnt{\small}
\SetAlCapFnt{\small}
\SetAlCapNameFnt{\small}
\SetAlCapHSkip{0pt}

\acmJournal{TOG}
\acmYear{2025} \acmVolume{44} \acmNumber{6} 
\acmArticle{232} 
\acmMonth{12}

\setcopyright{cc}
\setcctype{by}

\acmPrice{}
\acmDOI{10.1145/3763334}

\usepackage[dvipsnames,svgnames]{xcolor}
\usepackage[nomessages]{fp}

\newcommand{\maxnumR}{9.5}
\newlength{\maxlenR}
\newcommand{\barR}[2][red!25]{%
  \settowidth{\maxlenR}{\maxnumR}%
  \addtolength{\maxlenR}{\tabcolsep}%
  \FPeval\result{round(#2/\maxnumR:4)}%
  \FPeval\resultN{round(\result*50:0)}%
  \FPeval\resultN{min(100,\resultN)}%
  \FPeval\resultN{max(0,\resultN)}%
  \rlap{\color{RubineRed!\resultN}\hspace*{-.75\tabcolsep}\rule[-.05\ht\strutbox]{\result\maxlenR}{.95\ht\strutbox}}%
  \makebox[\dimexpr\maxlenR-\tabcolsep][r]{#2}%
}

\newcommand{\maxnumD}{4500}
\newlength{\maxlenD}
\newcommand{\barD}[2][green!25]{%
  \settowidth{\maxlenD}{\maxnumD}%
  \addtolength{\maxlenD}{\tabcolsep}%
  \FPeval\result{round(#2/\maxnumD:4)}%
  \FPeval\resultN{round(\result*75 - 25:0)}%
  \FPeval\resultN{min(100,\resultN)}%
  \FPeval\resultN{max(0,\resultN)}%
  \rlap{\color{green!\resultN}\hspace*{-.5\tabcolsep}\rule[-.05\ht\strutbox]{\result\maxlenD}{.95\ht\strutbox}}%
  \makebox[\dimexpr\maxlenD-\tabcolsep][r]{#2}%
}

\newcommand{\maxnumE}{0.30}
\newlength{\maxlenE}
\newcommand{\barE}[2][red!25]{%
  \settowidth{\maxlenE}{\maxnumE}%
  \addtolength{\maxlenE}{\tabcolsep}%
  \FPeval\result{round(#2/\maxnumE:4)}%
  \FPeval\resultN{round(\result*50:0)}%
  \FPeval\resultN{min(100,\resultN)}%
  \FPeval\resultN{max(0,\resultN)}%
  \rlap{\color{RubineRed!\resultN}\hspace*{-.5\tabcolsep}\rule[-.05\ht\strutbox]{\result\maxlenE}{.95\ht\strutbox}}%
  \makebox[\dimexpr\maxlenE-\tabcolsep][r]{#2}%
}

\newcommand{\maxnumC}{0.95}
\newlength{\maxlenC}
\newcommand{\barC}[2][red!25]{%
  \settowidth{\maxlenC}{\maxnumC}%
  \addtolength{\maxlenC}{\tabcolsep}%
  \FPeval\result{round(#2/\maxnumC:4)}%
  \FPeval\resultN{round(\result*50  + 10:0)}%
  \FPeval\resultN{min(100,\resultN)}%
  \FPeval\resultN{max(0,\resultN)}%
  \rlap{\color{RubineRed!\resultN}\hspace*{-.5\tabcolsep}\rule[-.05\ht\strutbox]{\result\maxlenC}{.95\ht\strutbox}}%
  \makebox[\dimexpr\maxlenC-\tabcolsep][r]{#2}%
}

\begin{document}
\title[Environment-aware Motion Matching]{Environment-aware Motion Matching}

\author{Jose Luis Ponton}
\orcid{0000-0001-6576-4528}
\affiliation{%
  \institution{Universitat Politècnica de Catalunya}
  \city{Barcelona}
  \country{Spain}
  \postcode{08034}
}
\email{jose.luis.ponton@upc.edu}

\author{Sheldon Andrews}
\orcid{0000-0001-9776-117X}
\affiliation{%
  \institution{École de technologie supérieure (ÉTS)}
  \city{Montreal}
  \country{Canada}
  \postcode{H3C 1K3}
}
\email{sheldon.andrews@gmail.com}

\author{Carlos Andujar}
\orcid{0000-0002-8480-4713}
\affiliation{%
  \institution{Universitat Politècnica de Catalunya}
  \city{Barcelona}
  \country{Spain}
  \postcode{08034}
}
\email{andujar@cs.upc.edu}

\author{Nuria Pelechano}
\orcid{0000-0002-1437-245X}
\affiliation{%
  \institution{Universitat Politècnica de Catalunya}
  \city{Barcelona}
  \country{Spain}
  \postcode{08034}
}
\email{npelechano@cs.upc.edu}

\renewcommand\shortauthors{Ponton~et al.}

\begin{abstract}
Interactive applications demand believable characters that respond naturally to dynamic environments. 
Traditional character animation techniques often struggle to handle arbitrary situations, leading to a growing trend of dynamically selecting motion-captured animations based on predefined features.
While Motion Matching has proven effective for locomotion by aligning to target trajectories, animating environment interactions and crowd behaviors remains challenging due to the need to consider surrounding elements.
Existing approaches often involve manual setup or lack the naturalism of motion capture. Furthermore, in crowd animation, body animation is frequently treated as a separate process from trajectory planning, leading to inconsistencies between body pose and root motion.
To address these limitations, we present \emph{Environment-aware Motion Matching}, a novel real-time system for full-body character animation that dynamically adapts to obstacles and other agents, emphasizing the bidirectional relationship between pose and trajectory.
In a preprocessing step, we extract shape, pose, and trajectory features from a motion capture database. At runtime, we perform an efficient search that matches user input and current pose while penalizing collisions with a dynamic environment. Our method allows characters to naturally adjust their pose and trajectory to navigate crowded scenes. 
\end{abstract}

%
%

%
%


\begin{teaserfigure}
\centering
 \includegraphics[width=1\linewidth]{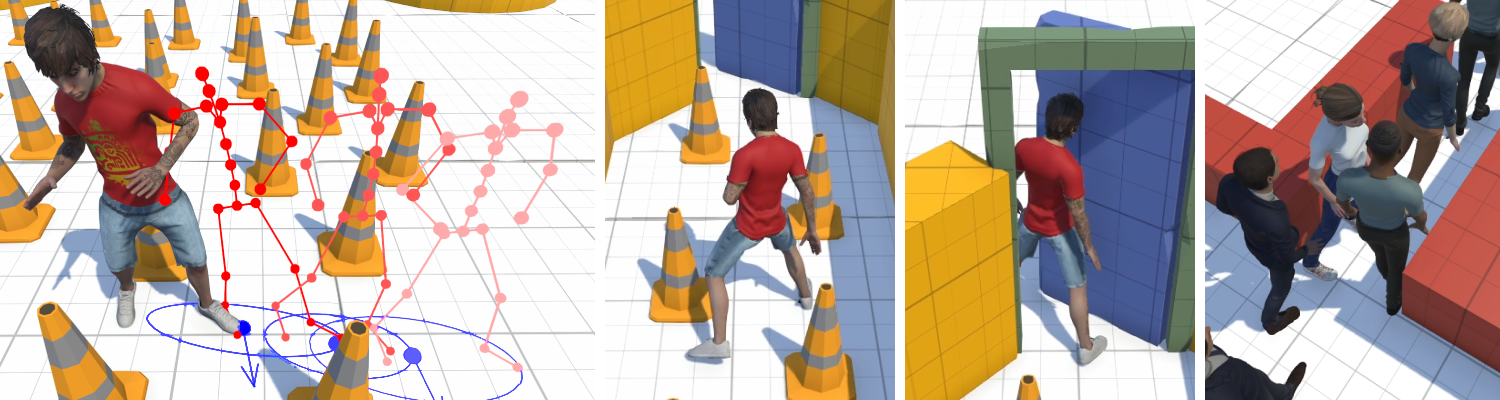}
 \caption{Environment-aware Motion Matching. Our real-time system enables characters to dynamically adapt their full-body pose and trajectory to navigate complex environments and interact with obstacles and other agents, seamlessly blending motion capture data with environmental constraints.}
\label{fig:teaser}
\end{teaserfigure}

\maketitle

\section{Introduction}
\label{sec:introduction}

Real-time interactive applications, such as video games, XR experiences, and virtual environments, increasingly demand dynamic and realistic character behavior in a wide range of different situations. The growing number of potential scenarios results in traditional animation, such as state-graph machines, being impractical. Consequently, leveraging motion-captured (mocap) animations to select poses based on contextual information \citep{holden2018gdc} has gained significant attention.

A prevalent technique for animating character locomotion is Motion Matching \citep{buttner2015motion, clavet2016}. This method searches extensive mocap databases for sequences of poses that align with a desired trajectory while ensuring smooth transitions from the current pose.

However, modeling interactions with the environment, between characters, or within crowds presents significant challenges. The number of relevant features to match expands rapidly, as it becomes necessary to consider environmental features in addition to the target trajectory and the current pose. Common approaches for environment interaction involve manually predefining markers within the scene that trigger specific animation clips \citep{allen2021, harrower2018gdc} depending on the character's location. This method necessitates character pre-alignment and offers limited flexibility and control during motion execution. Furthermore, the manual creation of these trigger points is a labor-intensive process. Alternative methods employing procedural animations, often relying on inverse kinematics \citep{alvarado2022generating}, frequently produce results that lack the naturalism inherent in mocap data.

Similar challenges arise when animating multi-character or crowd interactions. Notably, much of the existing research on crowd simulation treats body animation as an isolated process, separated from trajectory planning and collision avoidance \citep{hoyet2016perceptual, ferreira2024deformable, itatani:2024}. This decoupling often results in animations in which the body motion does not consistently align with the character's root movement, leading to foot sliding among other artifacts. In contrast, human interaction is typically a bidirectional process: the body pose influences the choice of trajectory, and conversely, the environment and constraints of a potential trajectory can dictate our body posture (e.g., choosing to sidestep in a narrow corridor).

In this paper, we introduce Environment-aware Motion Matching, a real-time system designed to dynamically animate the full-body pose of a character in response to its surroundings, taking into account nearby dynamic obstacles and other agents.
Unlike traditional methods that often decouple planning from animation, our system utilizes motion capture data to simultaneously compute both the pose and the trajectory of the agent, inherently incorporating collision avoidance. Our characters exhibit a natural adaptation to environmental elements by concurrently adjusting body pose and trajectory as needed.

During a preprocessing stage, we approximate the character's shape using 2D ellipses. This simple yet effective collision proxy ensures an accurate projection of the character's pose onto the floor, which is crucial for identifying optimal pose sequences to navigate narrow spaces or crowded areas. 

At runtime, we periodically perform a two-step search within a motion capture database. This process retrieves motion that adheres to the user's input while dynamically adapting to the environment. The integration of a simple collision proxy directly within the Motion Matching algorithm allows for natural mode changes to react to proximity constraints\textemdash both static and dynamic\textemdash imposed by the environment, enabling a flexible and context-aware character behavior.

The main contributions of our paper can be summarized as follows:
\begin{itemize}
    \item We introduce \textbf{Environment-aware Motion Matching}, a novel real-time framework that integrates a simple collision proxy directly into the Motion Matching algorithm. This enables natural, coupled reactions to static and dynamic environmental constraints, generating perfectly aligned body poses and root motion. This fundamentally addresses the prevalent issue of decoupled local steering and animation in crowd simulations, allowing for complex behaviors from simple user inputs.
    \item Our system is designed for straightforward integration into existing Motion Matching pipelines and supported by extensive optimizations ensuring real-time performance with minimal overhead. It simplifies data acquisition by requiring only a single-actor animation database, eliminating the need for multi-character captures or object labeling.
    \item We propose a flexible and extensible abstraction using novel environment features and obstacle penalizations. Unlike standard Motion Matching's direct query matching, our environment features are used to dynamically compute penalization factors, allowing for context-aware pose selection. This design facilitates easy extendability, enabling straightforward incorporation of diverse interaction types (e.g., height features).
\end{itemize}
\section{Related Work}
\label{sec:related_work}

In this section, we review the literature most relevant to our work. We begin by covering Motion Matching, then discuss existing crowd simulation methods that integrate body motion with local steering algorithms, and finally, compare our approach to reinforcement learning-based environment-aware methods.

\subsection{Motion Matching}
Motion Matching, initially introduced by \citet{buttner2015motion} as a greedy approximation to Motion Fields \citep{Lee2010motionfields}, was significantly advanced for Ubisoft's \textit{For Honor} \citep{clavet2016}. This technique emerged to address the limitations of traditional motion graphs \citep{kovar:2002, arikan:2002, lee2002motiongraphs, safonova2007motiongraphs, allbeck_planning_2011}, which often suffered from complex, rigid construction and maintenance. In contrast, Motion Matching's strength lies in its elegant simplicity: it relies on a large, unstructured motion capture dataset, a well-designed feature set, and an efficient nearest-neighbor search. This allows for real-time, flexible pose transitions and facilitates rapid iteration in animation style.

Despite the recent prominence of deep learning-based methods \citep{PFNN, starke:2018, motionVAE:2020, starke2022}, Motion Matching remains an essential animation tool in many AAA game productions. Its fast iteration times and high animation quality often outweigh the challenges deep learning methods face, such as the \textit{averaging} effect where high-frequency details are lost due to input signal ambiguity.

Recent research has expanded Motion Matching into various domains. \citet{holden2020} explored learning compact neural controllers from extensive animation data. \citet{2022:ponton:mmvr} applied Motion Matching to animate virtual reality avatars driven by head-mounted displays. Beyond locomotion, Motion Matching has been used for controllable gesture synthesis from speech \citep{motionmatching_gestures_2022} and for locomotion authoring based on footstep and gait cycle matching \citep{kim2024interactive}. Generative motion synthesis methods based on Motion Matching have also been proposed \citep{weiyu_2023_generativeMM}.

While Motion Matching has diversified in applications, most existing work primarily focuses on modifying query features or refining the standard nearest-neighbor search. Our approach, however, introduces novel environment features and a distinct two-stage search process. This is critical for adapting to dynamically changing environments and avoiding the necessity of prohibitively large, pre-annotated datasets.

\subsection{Crowd Simulation}
We demonstrate our approach in a crowd simulation setting, where methods tend to focus on solving the local steering of characters to move within nearby points while avoiding collisions with obstacles and other agents \citep{Helbing2000, Reynolds1999GDC, Pelechano2007, snape2011hybrid}.
For agents to move across large environments, a global pathfinding algorithm typically computes a sequence of waypoints, treating the environment as if empty. These waypoints then serve as attractors for local steering, which handles collision avoidance with other moving agents.
Finally, an independent animation layer is subsequently added to synthesize locomotion sequences that follow the input trajectory.
For example, \citet{sung2007continuous} employs a high-level path planning algorithm for global trajectories, with a motion graph then filling in animation sequences to match this global path. Similarly, \citet{yao2022CrowdSimulationDetailedBodyMotion} utilizes pathfinding and a social forces system for crowd simulation, animating characters via state machines for fixed interactions and PFNN \citep{PFNN} for locomotion. However, their agents lack the ability to adapt animations based on other agents' proximity or to adjust trajectories in response to current body animation.

Some methods propose long-term planning strategies to find smooth, collision-free trajectories with correctly aligned animations.
This includes searching small animation databases for complete solutions from start to goal positions \citep{lau2005behavior}. The navigation problem can also be decomposed into multi-domain problems \citep{Kapadia2013Multidomain}, where paths are first found on a navigation mesh and then refined in more complex domains that account for time and animations.
Global planning for collision-free trajectories can also be formulated as a non-linear optimization problem in 3D space-time, solving for agents as \textit{rods} via a quasi-Newton interior point solver \citep{modi2023multiagent}.

Managing inter-agent collisions in high-density crowds is a critical challenge, leading to the exploration of more accurate body representations.
\citet{hoyet2016perceptual} evaluated the importance of torso rotation in a simulated crowd to perceptually hide existing collisions.
To better model body shape and allow for motions like sidestepping, \citet{best2016real} and \citet{stuvel2016torso} employ 2D ellipses and 2D capsules, respectively, with RVO models, and more recently, \citet{ferreira2024deformable} utilizes deformable 2D ellipses within social forces frameworks.
For dense crowd collision resolution, \citet{gomez2024resolving} models the problem as an energy minimization for the entire crowd, focusing on correcting collisions for agents in place, but does not support adapting trajectories based on the environment.
Crucially, while these methods enhance collision handling and body representation, they typically maintain a separation between local avoidance mechanisms and the animation synthesis process, contrasting with our integrated approach.

\subsection{Reinforcement learning approaches}
Reinforcement learning (RL) has been successfully applied to character animation (see \citep{li2021survey, ariel2022ca-rl-survey} for recent surveys). Trained agents are capable of taking complex motor actions that are oriented toward accomplishing some (potentially high-level, long-horizon) task or achieving a goal based on a specified reward function. A distinguishing advantage of RL approaches is that motion skills, which encompass both skeletal animation and motion planning, are learned through experimentation. Strictly speaking, RL does not require training data; however, recent approaches do use motion capture data, either in the form of motion clips for kinematic motion synthesis~\citep{lee2006precomputing, Lee2010motionfields} or as reference data of natural human motion for physics-based approaches~\citep{peng2018deepmimic, peng2021amp, peng2022ase}. 
Although recent deep RL models provide plausible motion synthesis~\citep{peng2022ase, dou2023conditionalase, deng2024motiontransitions, xu2025parc}, training physics-based controllers to reproduce complex physical interactions remains an obstacle, requiring hours or days~\citep{ariel2022ca-rl-survey}. Even methods for efficient policy adaptation require significant resources and fine-tuning~\cite{adaptnet}, and challenges like catastrophic forgetting and mode collapse remain significant hurdles.
Furthermore, approaches based on generative adversarial imitation learning~\citep{peng2021amp, peng2022ase} struggle to reproduce exact animations from the reference dataset, which is an advantage of motion matching. There are also practical challenges, since physics-based character controllers are highly dependent on the design of the character model, e.g., joint and torque limits. 

These limitations contrast with the fast improve-and-test iterations supported by our approach, which is designed for precise control over trajectories and style, allowing animators to use the exact motions they capture. This is a key reason for its widespread adoption in AAA games, in addition to the facility of integrating it into game engines and crowd simulators. To our knowledge, no established RL-based method specifically targets navigation in tight, constrained spaces for a direct baseline. These fundamental differences in goals and technical approaches highlight why our method offers a faster, more controllable, and more artist-friendly workflow, complementing RL in the broader field of character animation.
\begin{figure*}[ht]
    \includegraphics[width=1.0\linewidth]{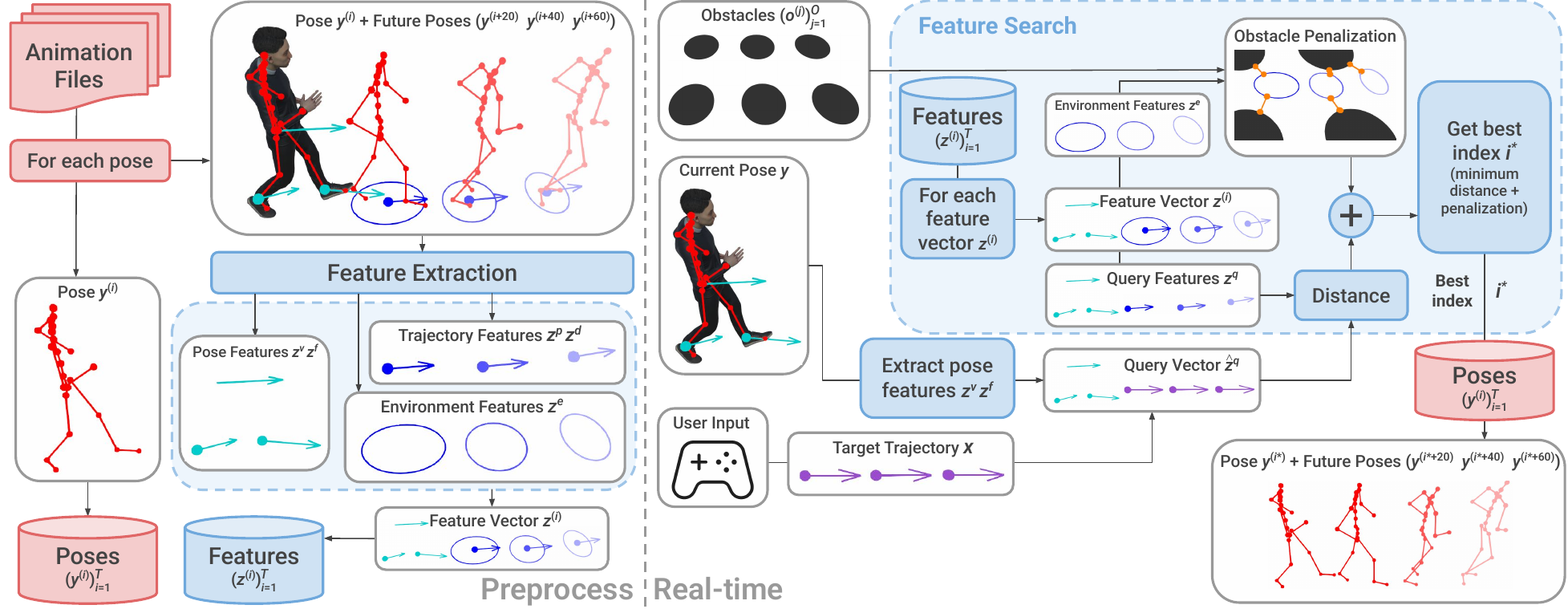}
    \caption{Overview of our Environment-aware Motion Matching pipeline. The system operates in two distinct stages: a preprocessing phase (Section~\ref{sec:method:preprocess}) and a real-time controller (Section~\ref{sec:method:controller}). The real-time controller utilizes user input and the current character pose to construct a query vector, which is then compared against the query features. Simultaneously, environment features guide the search by computing dynamic obstacle penalizations.}
    \label{fig:pipeline}
\end{figure*}

\section{Preliminaries: animation database, real-time user input, and environment constraints}
\label{sec:method:problem_definition}

Our animation database consists of an ordered sequence of\, $T$~poses $( \bm{y}^{(i)} )^{T}_{i=1}$. These poses are defined based on a humanoid skeleton with $J=24$ joints. The orientation of each joint is described relative to the local frame of its parent joint. Although the root joint is typically located at the hip, we introduce an additional bone to define the character reference frame. This is achieved by projecting the hip joint onto the ground plane and making the hip joint the child of this newly created root. The position of the newly added joint can be smoothed to minimize sudden changes. We refer to the 2D position of this projected hip joint on the ground plane as the character position. Additionally, we store the 3D local position of the hip joint within the character's local space. Consequently, the resulting pose vector is $\bm{y}^{(i)} \in \mathbb{R}^{4J + 3 + 2}$. Orientations are defined as quaternions. With this representation, we can accurately reproduce any part of the original animation by sequentially accessing poses $( \bm{y}^{(i)} \ \bm{y}^{(i+1)} \ \bm{y}^{(i+2)} \ \dots )$ given an initial pose vector.

The target trajectory $\bm{x} \in \mathbb{R}^{12}$ is defined similarly to several Motion Matching implementations \citep{clavet2016, holden2020}, where $\bm{x} = ( \bm{x^p} \ \bm{x^d} )$. Here, $\bm{x^p} \in \mathbb{R}^6$ and $\bm{x^d} \in \mathbb{R}^6$, represent the predicted 2D character positions and directions at 20, 40 and 60 frames in the future (assuming the application runs at 60\,Hz). As we shall see, these lookahead positions allow our characters to better adapt their animation to potential obstacles, thus minimizing the risk of collision. This input formulation provides sufficient flexibility to accommodate various input modalities for driving our system. For example, we can get target velocities from keyboard or game controller input and extrapolate future positions and directions using linear or more advanced methods, e.g., spring-based models. Alternatively, we can generate target trajectories from predefined paths or utilize scripted target velocities as produced by crowd simulation frameworks. 

The environmental constraints are defined as a list of $O$ obstacles $( \bm{o}^{(j)} )^O_{j=1}$, where each obstacle $\bm{o}^{(j)} \in \mathbb{R}^{d}$ has a dimensionality $d$ that depends on the obstacle type. Our framework is designed to accommodate various obstacle representations; however, in this work, we primarily focus on disks and ellipses. A disk ($d=3$) is defined by its 2D projected global center position and its radius. An ellipse ($d=5$) is defined by the 2D position of its center, the 2D semi-major axis vector, and the magnitude of the semi-minor axis (its direction can be determined by finding a vector perpendicular to the semi-major axis). The semi-major axis vector is aligned with the character's movement direction. When the character is stationary, we orient it using the character's forward vector.

Our objective is to periodically select poses $\bm{y}^{(i)}$ from the animation database for sequential playback, ensuring that the resulting motion adheres to both the user's input control and the environmental constraints. 

\section{Environment-aware Motion Matching}
\label{sec:method}

Our method comprises two distinct stages, as illustrated in Figure~\ref{fig:pipeline}: a preprocessing phase (Section~\ref{sec:method:preprocess}), and a real-time controller (Section~\ref{sec:method:controller}). The real-time controller uses the features extracted during the preprocessing stage to identify the most suitable pose based on the target trajectory, the character's current pose, and the surrounding environment.

\subsection{Feature Representation and Extraction}
\label{sec:method:preprocess}

This section describes the preprocessing stage, focusing on the motion features and their extraction from the animation database $(\bm{y}^{(i)})^T_{i=1}$.

From each pose $\bm{y}^{(i)}$, we extract three categories of features: pose features, trajectory features, and environment features. Pose and trajectory features are defined similarly to those used in existing Motion Matching implementations \citep{clavet2016, holden2020}. 
We introduce environment features that enable an environment-aware pose search.
We collectively refer to pose and trajectory features as query features, as they will be compared against a similarly defined query vector that encapsulates our target feature values.
In contrast, environment features are not directly compared to a predefined query vector. Instead, they are used to compute penalization factors based on a dynamic analysis of the surrounding scene. 
We believe that this distinction between query and environment features offers a valuable abstraction for the system's implementation, facilitating the addition or removal of features as required, as will be demonstrated later with the environment features.

\subsubsection{Query Features}
Query features are composed of pose and trajectory features.
Pose features are employed to minimize pose discontinuities during non-sequential transitions, i.e., when choosing the next pose from a different segment of the mocap database. We use the 3D linear velocities of the feet and hip joints, denoted as $\bm{z^v} \in \mathbb{R}^9$, and the 3D positions of the feet joints, $\bm{z^f} \in \mathbb{R}^6$. We selectively match these key joints to maintain a low-dimensional feature vector, thereby facilitating the search given the diverse nature of the features. Furthermore, excessively constraining the search space could significantly limit the number of viable transitions between poses.
Trajectory features are used to drive the motion according to the user's input. If only pose features were considered, our system would simply play back the animation database sequentially. Trajectory features enable the selection of poses that align with the movement intended by the user. Consistent with the target trajectory definition in Section~\ref{sec:method:problem_definition}, we use the 2D character positions $\bm{z^p} \in \mathbb{R}^6$ and directions $\bm{z^d} \in \mathbb{R}^6$ at 20, 40 and 60 frames into the future.
All features are expressed in character space and standardized to account for the varying magnitudes across different feature types. The complete vector of query features is defined as: $\bm{z^q} = ( \bm{z^v} \ \bm{z^f} \ \bm{z^p} \ \bm{z^d} ) \in \mathbb{R}^{27}$.

\subsubsection{Environment Features}
Environment features serve to dynamically evaluate the suitability of a given pose within a specific environmental context. In our work, we aim to define features that describe the character's overall body shape, enabling the automatic identification of pose sequences that can traverse a variety of obstacle configurations. For example, in a narrow corridor, a slight torso rotation might be necessary to pass another agent moving in the opposite direction.

We opted for a body representation that could be compactly integrated into the feature vector, thereby minimizing memory overhead, while simultaneously capturing different body orientations and shapes to readily distinguish motions such as sidestepping from forward walking. Although our framework can accommodate various types of representations depending on the specific requirements, we discarded disks due to their inability to differentiate between the aforementioned scenarios. More complex representations, such as convex hulls, were deemed unsuitable due to the increase in memory consumption and computational complexity. We selected ellipses to represent the 2D footprint of the body shape. Although the most compact representation, disks, can be described by a single real number (radius), ellipses require only two additional real numbers to describe the semi-major axis vector and the magnitude of the semi-minor axis, as detailed in Section~\ref{sec:method:problem_definition}. Similarly to the trajectory features, we define an ellipse for 20, 40, and 60 frames into the future. Consequently, we utilize the future 2D character positions stored in the trajectory feature $\bm{z^p}$ to position these ellipses. Thus, our principal environment features are defined as $\bm{z^e} \in \mathbb{R}^9$ (we experiment with additional environment features in Section~\ref{sec:eval:height-features}).

Ellipses are computed for each pose in the animation database by first determining the unit semi-major axis vector. This is achieved by normalizing the 2D displacement vector between the current character position and the subsequent one. The semi-minor axis vector is then computed by finding a perpendicular vector to the semi-major axis (by swapping its components and negating the first one). Finally, to determine the magnitude of both axes, all joints in the skeleton are projected onto these two vectors, and the maximum projected distance is used. The resulting ellipse is stored in character space.

It is important to note that while we utilize ellipses to represent the body shape, our environment features framework is designed for a wide variety of body representations. This offers the flexibility to use more complex representations if higher accuracy is needed, as the system's core abstraction of computing penalizations remains the same. For instance, as discussed in Section~\ref{sec:eval:height-features}, we employ height features to determine whether a character should jump or crouch to avoid vertical obstacles. This would involve adding two real numbers per ellipse, representing the minimum and maximum vertical components in character space.

\subsubsection{Feature Vector and Feature Database}
The final complete feature vector $\bm{z}$ is the concatenation of the query features $\bm{z^q}$ and the environment features $\bm{z^e}$:
\begin{equation}
    \bm{z} = \begin{pmatrix} \bm{z^q} \ \bm{z^e} \end{pmatrix} \in \mathbb{R}^{36} \,.
\end{equation}
Thus, the feature database $( \bm{z}^{(i)} )^{T}_{i=1}$ has a corresponding entry $\bm{z}^{(i)}$ for each pose $\bm{y}^{(i)}$ in the animation database $( \bm{y}^{(i)} )^{T}_{i=1}$.

\subsection{Feature Search}
\label{sec:method:search}

Having constructed the feature database $( \bm{z}^{(i)} )^{T}_{i=1}$ and the corresponding animation database $( \bm{y}^{(i)} )^{T}_{i=1}$, our objective, given the character's current pose $\bm{y}$ and the target trajectory $\bm{x}$ (as defined in Section~\ref{sec:method:problem_definition}), is to identify the optimal matching pose $\bm{y}^{(i^*)}$. The subsequent poses in the animation sequence $( \bm{y}^{(i^*+20)} \ \bm{y}^{(i^*+40)} \ \bm{y}^{(i^*+60)} )$, should align with the current pose and target trajectory while avoiding collisions with environmental obstacles $( \bm{o}^{(j)} )^{O}_{j=1}$.

The feature search process is conceptually divided into two main steps. First, we construct a query vector $\hat{\bm{z}}^{\bm{q}}$ containing the target values for our query features. Second, the environment features, for which direct targets are not defined, are used to compute penalization factors that influence the selection process.

We begin by creating the query vector $\hat{\bm{z}}^{\bm{q}}$ that contains the target query features. The pose features $\bm{z^v}$ and $\bm{z^f}$ are directly derived from the current character pose $\bm{y}$. The trajectory features $\bm{z^p}$ and $\bm{z^d}$ are constructed based on the provided input $\bm{x}$, as detailed in Section~\ref{sec:method:problem_definition}. 

Subsequently, we initiate the search by iterating through all feature vectors $\bm{z}^{(i)}$ in the feature database $( \bm{z}^{(i)} )^{T}_{i=1}$. For each feature vector, we perform the following steps sequentially (optimizations of these steps are discussed in Section~\ref{sec:method:optimizations}):

\paragraph{(1) Query features distance} We compute the Euclidean distance between the query vector $\hat{\bm{z}}^{\bm{q}}$ and the query features $\bm{z^q}$. Since all query features are standardized, we employ weights $(\lambda^v \ \lambda^f \ \lambda^p \ \lambda^d)$ to control the relative importance of each feature type. Typically, these weights are set to 1.

\paragraph{(2) Obstacle penalizations} For each of the three future ellipses associated with the environment features $\bm{z^e}$, we determine their global position based on the future 2D character positions $\bm{z^p}$ and then iterate through all nearby obstacles to calculate the minimum distance between the ellipse and each obstacle.

\begin{figure}
    \centering
    \includegraphics[width=1\linewidth]{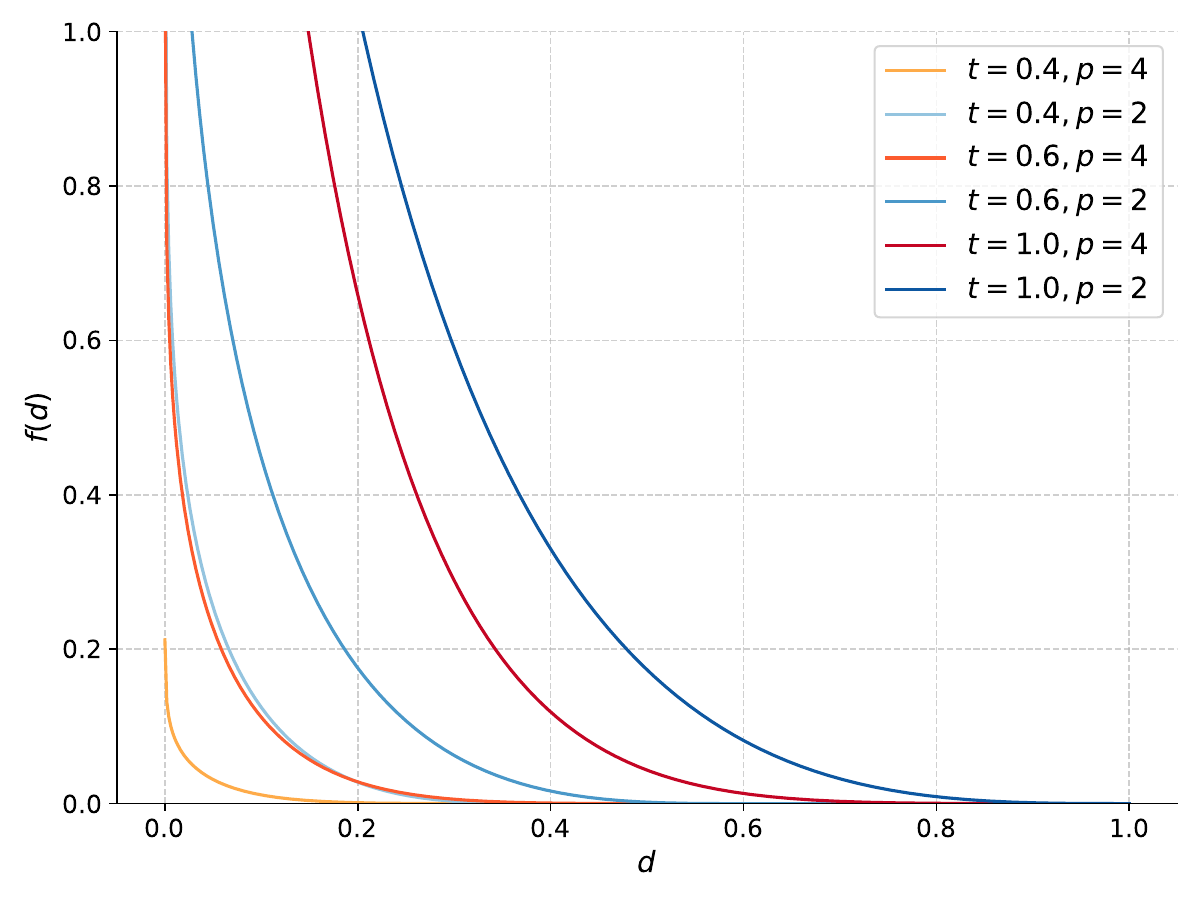}
    \caption{The log-barrier function used for obstacle penalization. This plot illustrates the penalty $f(d) = -(t - d)^p \log \left( \frac{d}{t} \right)$ as a function of the distance $d$ to an obstacle. The penalty is zero when $d \ge t$ (the obstacle distance threshold), and increases exponentially as $d$ approaches zero. The parameter $p$ controls the steepness of this exponential growth, ensuring a strong repulsion as the character gets close to an obstacle.}
    \label{fig:log-barrier}
\end{figure}

\paragraph{(3) Log-barrier function} Each computed distance $d$ is then evaluated using a log-barrier function (illustrated in Figure~\ref{fig:log-barrier}), which imposes an exponentially increasing penalty as the distance approaches zero, such that
\begin{equation}
    f(d) =
    \begin{cases}
        -(t - d)^p \log \left( \frac{d}{t} \right) & \text{if} \ d < t \\
        0 & \text{otherwise}
    \end{cases}
\end{equation}
where $t$ represents the obstacle distance threshold, defining the range within which penalization begins, and $p$ controls the rate at which the penalty increases as the distance decreases. The log-barrier function yields near-zero penalization around $t$ and grows exponentially as $d$ approaches zero. This barrier is inspired by the Incremental Potential Contact method~\cite{Li2020IPC}.

\paragraph{(4) Weighted penalizations} To further modulate the influence of the penalization terms, we introduce two levels of weights. First, a general environment weight $\lambda^e$ is applied to all penalizations. Second, we employ additional weights $\lambda^{40}$ and $\lambda^{60}$ to adjust the significance of the penalizations arising from the second (40 frames) and third (60 frames) future ellipses, respectively. 
Ideally, we aim to identify a trajectory that satisfies all obstacle constraints. However, given the limited size of the animation database, achieving this may not always be feasible. In such scenarios, prioritizing the avoidance of immediate future collisions is often acceptable ($\lambda^{40}=0.4$ and $\lambda^{60}=0.1$).

\paragraph{(5) Final score} Finally, the weighted query distance and all weighted penalization values are summed to obtain a final score for the feature vector $\bm{z}^{(i)}$. We get the index $i^*$ corresponding to the feature vector with the lowest overall score. The corresponding pose $\bm{y}^{(i^*)}$ is then retrieved from the animation database for playback.

\section{Real-time Environment-aware Character Controller}
\label{sec:method:controller}

Our real-time environment-aware character controller uses the features and search mechanism detailed in Section~\ref{sec:method}. This section elaborates on how these concepts are integrated into our real-time control framework.

\subsection{Overview}
The real-time controller operates through two tasks executed at different frequencies:

\paragraph{Every $n$ frames}
The controller first gathers nearby obstacles. Subsequently, as described in Section~\ref{sec:method:search}, it constructs the query vector $\bm{\hat{z}^q}$ and performs a feature search to identify the optimal matching pose $\bm{y}^{(i^*)}$. The character's position and orientation in this best-matching pose define the animation space, while the current character's position and orientation define the character space. A transformation is then computed to enable the playback of the new pose within the current character space. Immediately following this search, the index $i^*$ is incremented by one, as detailed in the next step. In our experiments, we set $n=10$ for a 60\,Hz application.
    
\paragraph{Every frame}
The system advances the current best-matching pose by one frame in the animation database. Formally, let $c$ represent the number of frames elapsed since the last search. The character is animated with the pose $\bm{y}^{(i^*+c+1)}$. This assumes that the animation database was captured at the same frame rate as the application. In practice, all counters are time-based rather than frame-based, allowing for smooth animation playback even with variable frame rates. Any post-processing steps are also executed during this per-frame process. For example, we employ inertialization \citep{bollo2017inertialization} after each new search, thus every $n$ frames, to blend seamlessly between changes in pose.

\begin{figure}
    \centering
    \includegraphics[width=1\linewidth]{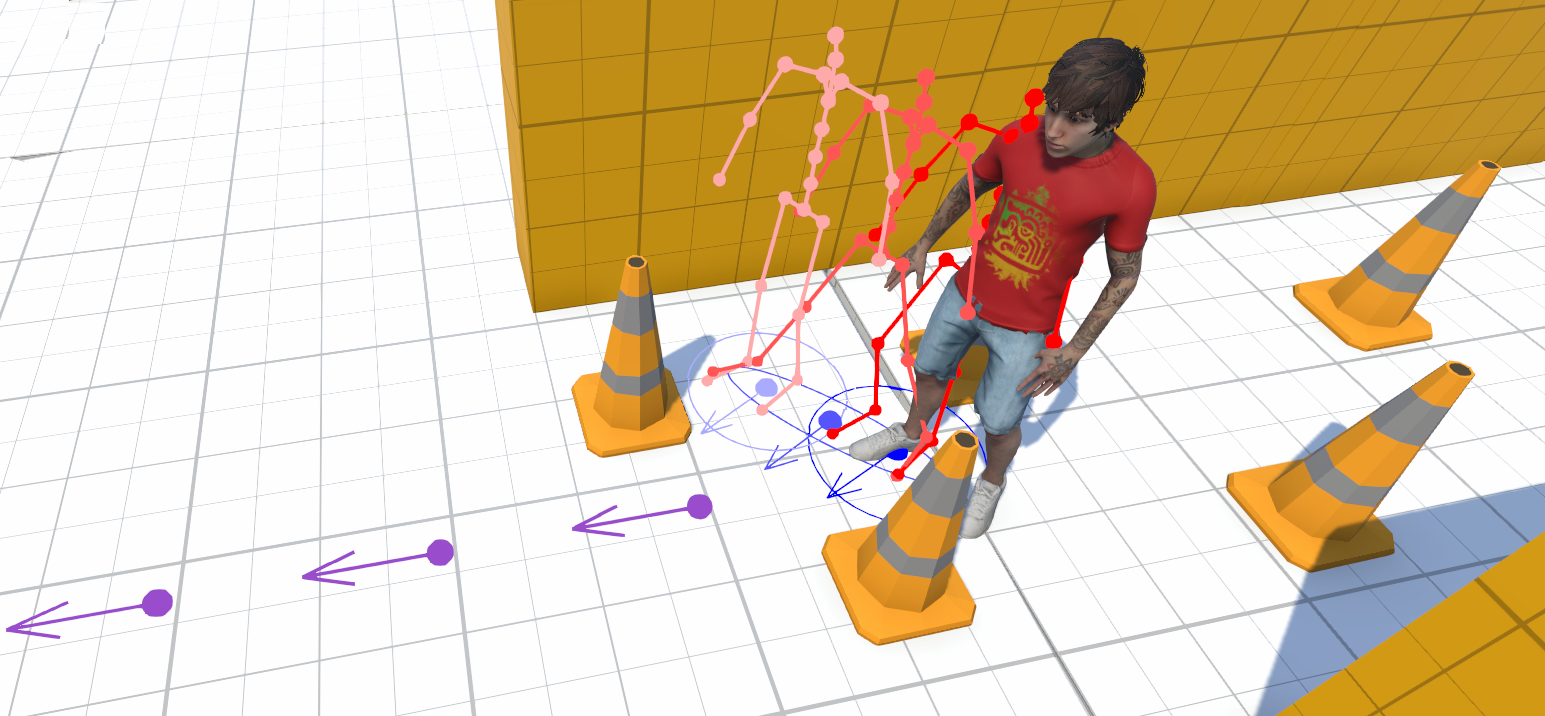}
    \caption{Purple arrows indicate the target trajectory derived from user input. Our method, while being aware of the cone obstacle, dynamically adjusts the character's path, causing it to step right to avoid collision. This demonstrates how the system integrates environmental constraints into root motion, ultimately enabling the character to reach the target trajectory while naturally avoiding obstacles.}
    \label{fig:target-trajectory-vs-ellipses}
\end{figure}

Collision avoidance implies that the character's final trajectory might not match the trajectory extracted from the provided input, since it has to be adjusted to the environment (see Figure~\ref{fig:target-trajectory-vs-ellipses}). Since our search algorithm handles the extraction of the most appropriate motion for each situation, the character's motion is entirely driven by the root motion embedded within the animation database, rather than the trajectory derived from the provided input.
Consequently, when generating a new target trajectory based on the provided input, the starting point is always considered to be the character's current position. For applications requiring more immediate responsiveness to input, it would be feasible to maintain a separate target position derived directly from the given input and subtly guide the character towards this target. However, this approach might introduce some visual artifacts, such as foot sliding, necessitating additional post-processing techniques like foot locking.

\subsection{Capture of the Animation Database}
A high-quality animation database is crucial for realistic and responsive control. Despite addressing environment interactions, our method only requires a single actor, without needing to track or tag obstacles during capture.

We first captured diverse locomotion data (similar to Motion Matching) for obstacle-free scenarios. Furthermore, we designed and captured various scenarios that involve both static and dynamic obstacles with a single actor. We aimed for broad animation coverage and iteratively added new situations whenever our character controller did not exhibit the desired behavior. Our capture diagrams, detailing the actor's instructions, are included in the supplemental material to facilitate reproducibility.

While comprehensive animation coverage is key, redundant clips of the same motion are avoided to prevent constant transitions during loops. To achieve this, our search process incorporates a slight reduction in the final score if the character continues the currently playing animation clip instead of transitioning to a new one.

Although a carefully curated animation database (high diversity and minimal redundancy) yields the best performance, we achieved high-quality results with minimal manual effort. For instance, in this work, we utilize raw animations captured using an Xsens Awinda mocap suit. By instructing the actor to follow our pre-designed navigation tasks, we can generate new, fully functional animation databases in approximately two hours (including preparation, capture, file exporting, and application import). This rapid iteration enables the easy capture of different movement styles, and simply by swapping the animation database, entirely different styles can be obtained (see accompanying video).

\subsection{Feature Weights}
Through our experiments, we have found that setting the query feature weights $( \lambda^v_0 \ \lambda^f_0 \ \lambda^p_0 \ \lambda^d_0 )$ to one (subscript 0 indicates the default value) and the environment weights to $\lambda^e_0=5$, $\lambda^{40}_0=0.4$, $\lambda^{60}_0=0.1$ generally yields good results. We also defined two parameters for the log-barrier function: the exponent $p$ is set to 4 for all experiments, while the threshold $t$ is set to 0.4 for navigating highly constrained environments and 0.6 for more general usage.

Furthermore, we define and provide reference values for two commonly used high-level controls in Motion Matching (Responsiveness and Continuity), along with two novel high-level controls (Evasion and Anticipation) which offer intuitive ways to adjust the behavior of the character controller without needing to directly manipulate the individual low-level feature weights:

\paragraph{\textit{Responsiveness} $\omega_r\in \mathbb{R}_{>0}$.} Default $\omega_r=0.1$. Modifies the trajectory feature weights to adjust the degree to which the system adheres to the target trajectory. Higher values will result in the character more closely following the intended trajectory:
\begin{align}
    \lambda^p = \lambda^p_0 \, \omega_r &&
    \lambda^d = \lambda^d_0 \, \omega_r
\end{align}

\paragraph{\textit{Continuity} $\omega_c\in \mathbb{R}_{>0}$.} Default $\omega_c=0.1$. Modifies the pose feature weights to control the emphasis on minimizing pose discontinuities. Increasing this weight can reduce the number of potential transitions within the animation database. Still, it may be beneficial in scenarios where maintaining continuity is critical (e.g., during airborne phases of a jump):
\begin{align}
    \lambda^v = \lambda^v_0 \, \omega_c &&
    \lambda^f = \lambda^f_0 \, \omega_c
\end{align}

\paragraph{\textit{Evasion} $\omega_e\in [0, 1]$.} Default $\omega_e=0.01$. Defines the minimum scaling factor applied to the direction weight when obstacles are nearby. 
Typically, when the user inputs a desired movement direction, we want the character's body orientation to adapt optimally to the surrounding obstacles based on our animation database. 
Therefore, we have found it beneficial to reduce the direction weight $\lambda^d$ when the character is close to obstacles.
Let $\bm{o}$ be the nearest obstacle and $\bm{p}$ the current position of the character. We linearly interpolate the direction weight towards a target weight defined as:
\begin{equation}
    \lambda^d =
    \begin{cases}
        \lambda^d_0 \, \max\left(\omega_e, \frac{|| \bm{o} - \bm{p} ||}{t} \right) & \text{if } || \bm{o} - \bm{p} || < t \\
        \lambda^d_0 & \text{otherwise}
    \end{cases}
\end{equation}
Note that in scenarios where precise target directions are provided manually and high control is required, this parameter can be set to~1, effectively disabling its influence.

\paragraph{\textit{Anticipation} $\omega_a\in \mathbb{R}_{>0}$.} Default $\omega_a=2.0$. Scales the environment features weight $\lambda^e$ based on the character's target speed $\dot{v}$. Trajectory feature weights $\lambda^p$ become more influential at higher speeds due to the greater distances in the generated trajectories. This increased influence can sometimes lead the character to disregard obstacles during fast movements like running. To counteract this, we modulate the environment weight as follows:
\begin{equation}
    \lambda^e = \max \left( \lambda^e_0 \cdot \dot{v} \cdot \omega_a, \lambda^e_0 \right)
\end{equation}
The higher the anticipation, the earlier the character will avoid obstacles without having to slow down. Nevertheless, we always keep a minimum base weight to avoid completely ignoring environment features when idle (essential to evade dynamic obstacles, see Section~\ref{sec:eval:dynamic-obstacles}).

The high-level parameters \textit{Evasion} and \textit{Anticipation} allow for fine-tuning the collision avoidance strategy that has been incorporated into the motion matching approach through the environment features.

\subsection{Optimizations}
\label{sec:method:optimizations}

A key consideration when integrating environment features is their potential performance impact on the search. Search operations involving only query features (i.e., ignoring obstacles) can benefit from accelerating data structures such as Bounding Volume Hierarchy (BVH), scaling logarithmically with the size of the animation database. However, environment features, which depend on the dynamic elements within the scene, require a linear traversal of the database, scaling with the number of poses and obstacles. Obstacle penalization is also more computationally demanding than simple vector distance checks. To achieve real-time performance, we next describe some critical optimizations that we implemented. A detailed evaluation and performance analysis of these optimization strategies is presented in Section~\ref{sec:evaluation:performance}.

\subsubsection{Base optimizations}
We first implemented a set of core optimizations that do not modify the algorithm's behavior and should always be employed to significantly accelerate the feature search.

A crucial optimization is the early rejection of feature vectors whose cumulative score exceeds that of the current best candidate (as our objective is to find the minimum score). The entire feature search is structured around this principle: we calculate the query feature distance first and proceed to obstacle penalization only if it is lower than the current best score. During penalization, for each obstacle, we early reject the feature vector as soon as the score is higher than the current best candidate. We evaluate obstacles sequentially for the three future ellipses (first, second, then third), leveraging the lower weights of the later ellipses ($\lambda^{40}=0.4$ and $\lambda^{60}=0.1$) for faster rejection based on the first ellipse's penalties. Within each ellipse, simpler obstacles (disks) are processed before more complex ones (ellipses).

To further enhance the efficiency of early feature vector rejection, we experimented with pre-processing trajectories based on their clearance, intending to prioritize the checking of less restricted trajectories. However, this approach did not yield significant performance improvements.

We also optimize the search by computing obstacle penalizations only for potential intersections. Before the feature search, we pre-select relevant obstacles if their center is within $r_{obs}+t+r_{ellipse}$ of the character's future trajectory positions $\bm{z^p}$, where $r_{obs}$ is the obstacle's longest radius, $t$ is the distance threshold, and $r_{ellipse}=0.9$\,m (about half the arm span). If there are no potential obstacles for any ellipse, we execute a BVH-accelerated search using only query features.

Finally, our method naturally supports a data-oriented design. Contiguous memory storage of feature vectors and sequential access optimize the CPU cache and simplify SIMD integration for substantial performance improvements.

\subsubsection{Temporal coherence}
While our base optimizations achieve real-time performance, further minimizing the animation engine's impact on the frame budget is desirable to match traditional Motion Matching speeds ($0.22 \pm 0.05$ms, Table~\ref{tab:performance:ablation}). 

Exhaustively checking every feature vector for obstacle penalization is too costly (see Section~\ref{sec:evaluation:performance}). We devise an optimization strategy that aligns with Motion Matching's principles: operating on large, unstructured animation arrays without complex graph structures and without limiting pose transitions (a prior issue with Motion Graphs \citep{kovar:2002, arikan:2002, yin:2005}).

We exploit temporal and local coherence. Typically, Motion Matching databases contain long, unstructured sequences with local pose similarity and continuity. We also observe that many poses within these databases represent less distinctive motion. For instance, idle sequences between actions, which are common in Motion Matching to streamline transitions, often contain prolonged periods of very similar poses. These can be effectively condensed by identifying key poses that adequately represent the entire segment, thereby reducing the number of feature vectors that need to be explicitly checked during the search. This consolidation allows for more efficient early skipping during the search process.

To take advantage of local coherence, we first introduce a minimum search stride (we use a stride $k$ of 8 frames, or 133 ms at 60 Hz). To accelerate searches through extended periods of similar poses, such as idle states, we also employ an adaptive threshold. In a preprocessing step, we identify and store indices of feature vectors that are at least $k$ frames apart and exhibit a minimum difference of 5\,\% (relative to each feature's range). This creates a sparse array of representative feature vectors.
During the search, we initially iterate through these representative indices. If a representative index yields a score better than the current best, we then perform a more exhaustive search by examining all feature vectors between the halfway points of the preceding and succeeding representative indices. This approach resembles a two-level hierarchical search. 
%

During the coarse search through representative feature vectors, we dynamically adjust the stride based on the dissimilarity between the current representative $\bm{z}^{(i)}$ and the current best feature vector $\bm{z}^{(i^*)}$ (with a total score of $s^*$). The dissimilarity is quantified by the Euclidean distance $s = ||\bm{z}^{(i)} - \bm{z}^{(i^*)}||$. This distance is then used to compute the stride $k'$ for the representative vector search using the following formula:
\begin{equation}
    k' = \max \left( 1, v \floor*{\sqrt{\frac{s}{s^*}}} \right)
\end{equation}
where the parameter $v$ controls the search aggressiveness; smaller values lead to more conservative searches with smaller strides, while larger values encourage larger strides, potentially skipping more representative vectors.

The underlying intuition is that early in the search, when a good initial best score $s^*$ has not yet been established, the stride $k'$ will tend towards 1, ensuring that the most representative vectors are examined. As the search progresses and $s^*$ decreases, if the current representative $\bm{z}^{(i^*)}$ is significantly different from the current best $\bm{z}^{(i^*)}$ (i.e., $s$ is large relative to $s^*$), the resulting $k'$ will be greater than 1, allowing the algorithm to skip over less promising representative vectors.
Conversely, when the current representative is very similar to the best one ($s \approx s^*$), the stride $k'$ will become 1, therefore, having a more careful examination of the neighboring feature vectors. 

Finally, we begin the search from a representative feature vector similar to the one previously selected. Since motion tends to be cyclic, this helps quickly establish a good initial best score $s^*$ and facilitates the early rejection mechanism.  More precisely, our feature search begins approximately 1\,\% (relative to the length of the representative feature vector array) before the best-matched index from the previous frame.
\section{Evaluation}
\label{sec:evaluation}

In this section, we provide an in-depth evaluation of Environment-aware Motion Matching across a variety of scenarios and configurations. Subsequently, we present an ablation analysis and comparative studies to justify our design choices, including comparisons with standard Motion Matching, an examination of our optimization strategies, and a discussion on disk-based versus ellipse-based body representations.

We implemented Environment-aware Motion Matching within the Unity game engine, utilizing data-oriented programming principles and the Burst compiler to optimize the search process. The system was evaluated on a PC equipped with an Intel Core i7-12700k CPU, 32GB of RAM, and an NVIDIA GeForce RTX 3090 GPU (used solely for rendering).

\begin{figure}
    \centering
    \includegraphics[width=1\linewidth]{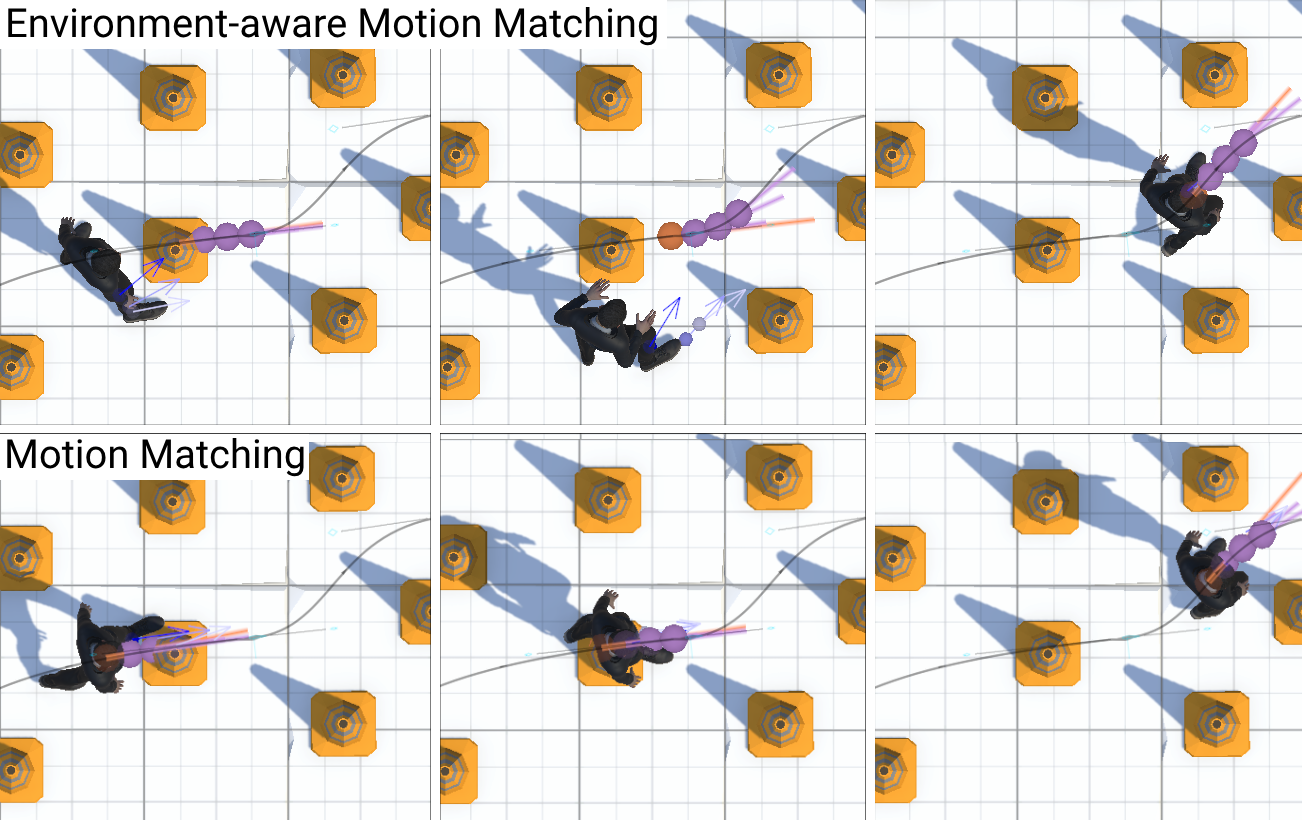}
    \caption{Comparison of our Environment-aware Motion Matching (top row) with standard Motion Matching (bottom row) in an obstacle-filled scene. The black line indicates the target path, and purple points denote future target positions. Blue points illustrate the system's selected future character positions. Our method (top) dynamically detours to avoid the central obstacle, showcasing its environment awareness and the coupled pose-trajectory selection. In contrast, standard Motion Matching (bottom) disregards the obstacle, leading to collisions. This highlights our system's bidirectional control, in contrast to typical decoupled animation pipelines.}
    \label{fig:obstacle_trajectory}
\end{figure}

\subsection{Results and Experiments}
\label{sec:evaluation:experiments}

In this section, we demonstrate the versatility of our method across a variety of scenarios and configurations. Each experiment is accompanied by an illustrative image and a segment of the accompanying video.

\subsubsection{Constrained Environment}
We designed a virtual scene incorporating a variety of obstacles to evaluate our method's ability to naturally adapt to diverse situations.

\begin{figure*}
    \centering
    \includegraphics[width=1\linewidth]{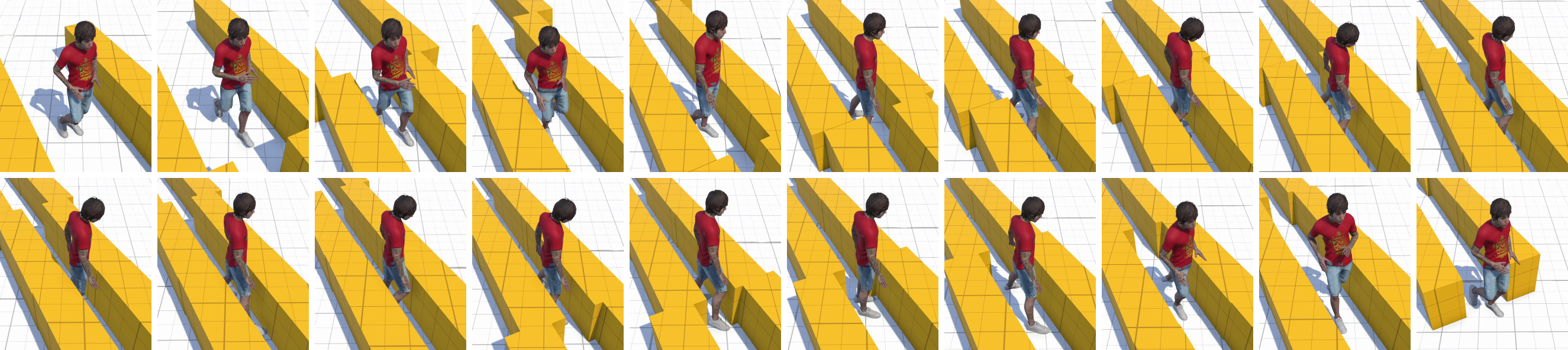}
    \caption{Character navigating a progressively narrowing corridor. The top sequence shows the character running in the wider sections. As the corridor narrows to 0.35 meters, the character reduces speed and transitions to side-stepping. When the corridor becomes wider again, the system switches back to a running animation. This demonstrates the system's ability to adapt body pose and speed to tight environmental constraints.}
    \label{fig:super_narrow_corridor}
\end{figure*}

First, we created a corridor that progressively narrows, eventually reaching a width of only 0.35 meters.
Figure~\ref{fig:super_narrow_corridor} shows the character's behavior: it initially runs unconstrained in the widest section. As the corridor narrows, the character reduces its velocity and then begins to turn its body, ultimately traversing the narrowest section by sidestepping. Once the corridor widens again, the character naturally resumes running.

\begin{figure*}
    \centering
    \includegraphics[width=1\linewidth]{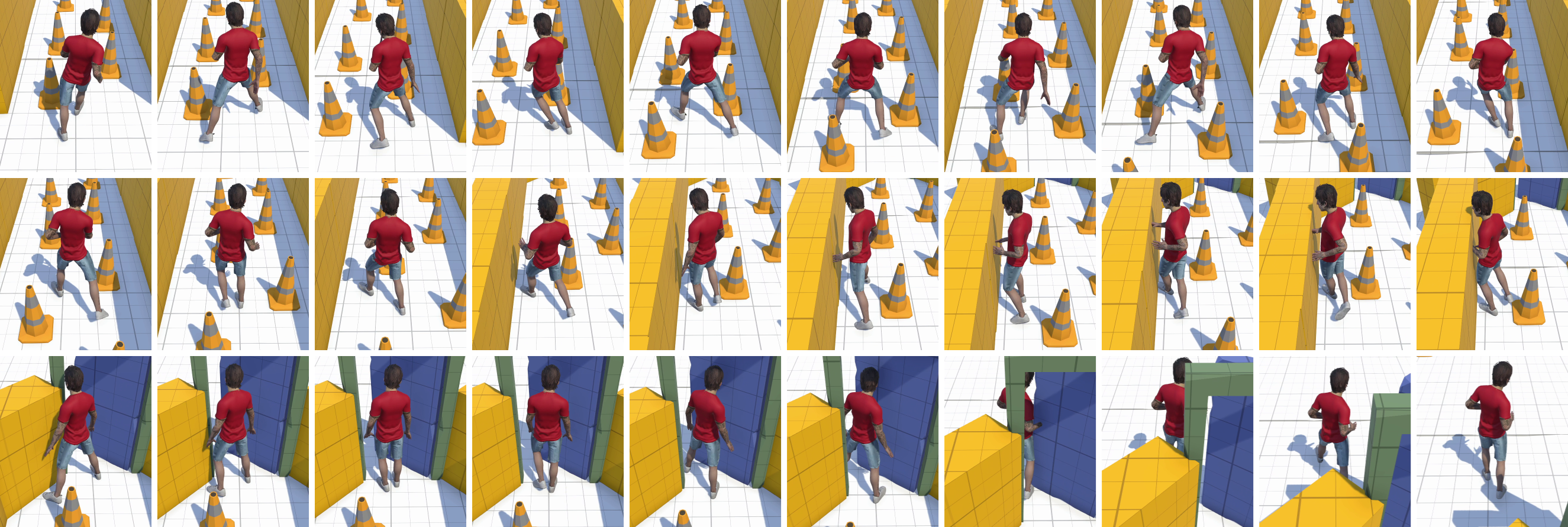}
    \caption{Character traversing a corridor with zigzag cone obstacles. Despite receiving only a continuous forward input, our system enables the character to dynamically zigzag, maintain velocity, and transition to side-stepping through narrower sections or partially closed doorways. This illustrates the generation of complex trajectories from simple user commands due to environment awareness.}
    \label{fig:zigzag_door}
\end{figure*}

Second, we designed a corridor containing cone obstacles arranged in a zigzag pattern. Figure~\ref{fig:zigzag_door} demonstrates how the character navigates this setup. Initially it zigzags to avoid the cones while maintaining velocity.
At a certain point, it opts to sidestep until reaching a nearly closed door (note that the cones are not perfectly evenly distributed). The character then traverses this partially obstructed doorway by sidestepping. 
It is crucial to highlight that throughout the duration of this corridor traversal, \textbf{the only input received by the system is a continuous forward command} (e.g., pressing the \textit{up} arrow key). Despite this simple linear target trajectory, our system generates rich and complex motions, such as zigzagging and carefully passing through the narrow opening.

\begin{figure}
    \centering
    \includegraphics[width=1\linewidth]{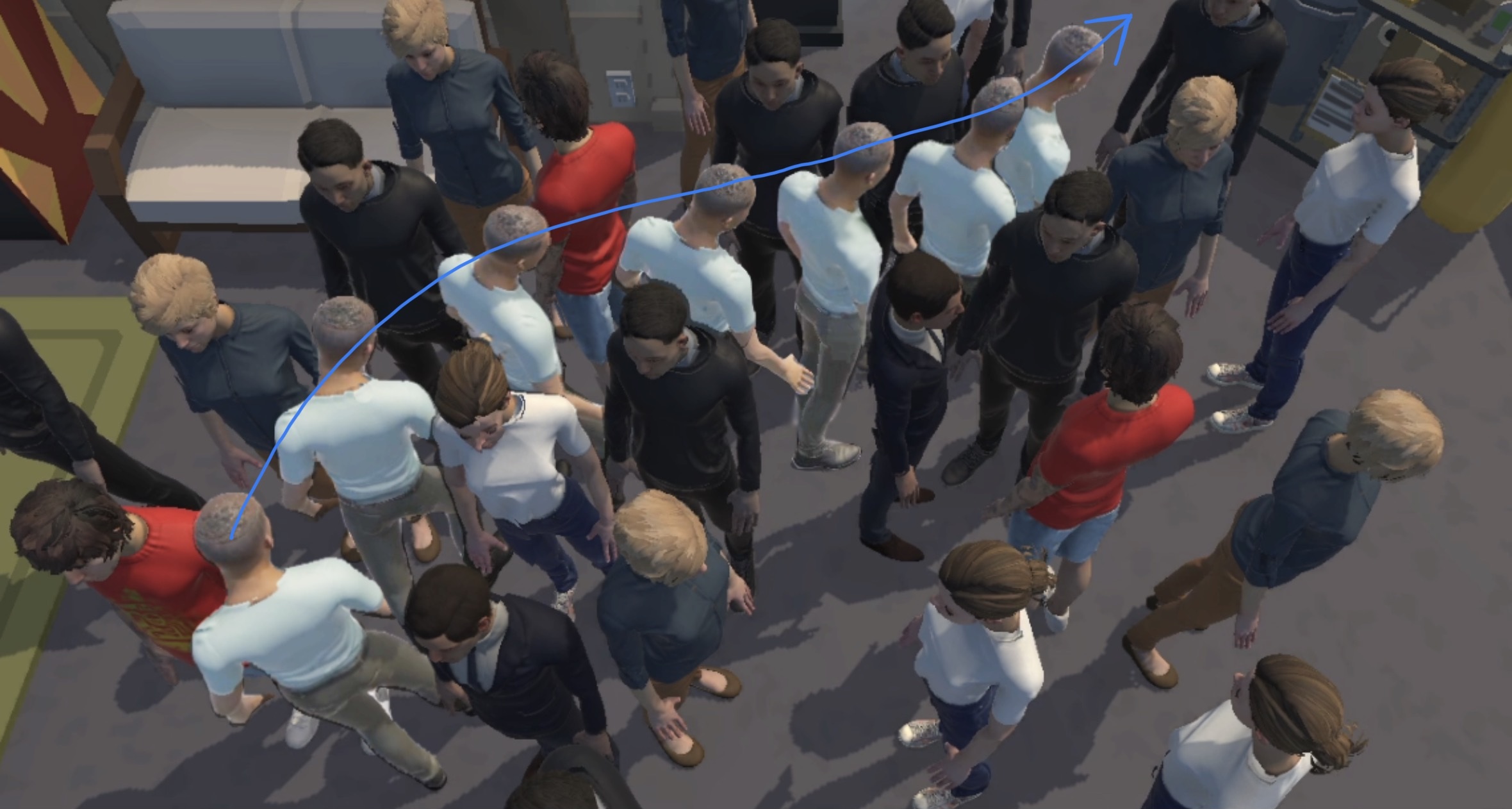}
    \caption{Character navigating a high-density crowd. The ligth blue t-shirt character walks slowly and carefully between other stationary characters, demonstrating the system's ability to find and adapt to paths within extremely confined, high-density environments.}
    \label{fig:high-density}
\end{figure}

Finally, Figure~\ref{fig:high-density} showcases our method in a high-density crowd scenario, where other characters are treated as static obstacles. The main character carefully navigates by slowly moving between them, demonstrating its ability to find paths and adapt its body in extremely tight spaces.

\subsubsection{Dynamic Obstacles}
\label{sec:eval:dynamic-obstacles}
A key advantage of our environment features is their capability to compute real-time penalizations without making assumptions about the environment. This means that our system can quickly adapt to dynamic obstacles. We approximate obstacles using simple proxy primitives, such as disks, with their positions aligned with the rendered objects. For dynamic objects, we also include additional proxies placed according to the obstacle's expected future position (ideally matching the time horizon of our agent's trajectory and environment features).

\begin{figure*}
    \centering
    \includegraphics[width=1\linewidth]{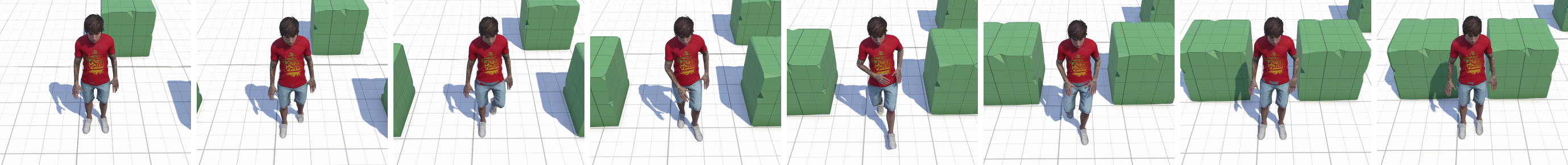}
    \caption{Character adapting to two moving cubes with no user input. The character, initially stationary, automatically takes a few steps forward to avoid colliding with the approaching cubes. This demonstrates the system's real-time adaptation to dynamic obstacles without user intervention.}
    \label{fig:moving_cubes}
\end{figure*}

Figure~\ref{fig:moving_cubes} shows a scenario in which two moving cubes approach a stationary character. \textbf{Without any user input}, our system automatically causes the character to take a few steps forward to avoid collision as the cubes approach.

\begin{figure*}
    \centering
    \includegraphics[width=1\linewidth]{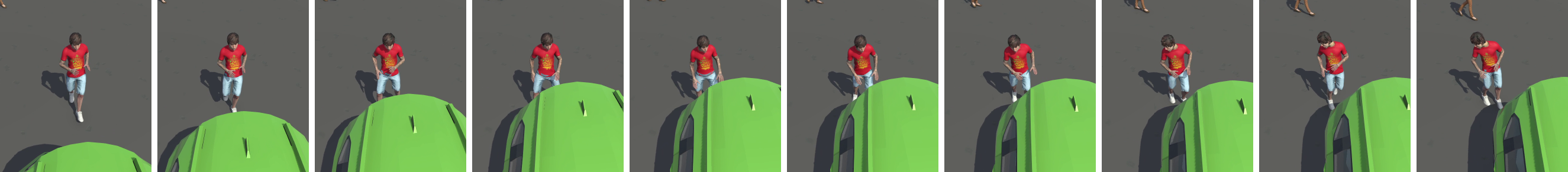}
    \caption{Character avoiding a moving car despite conflicting user input. The user inputs a forward movement (towards the car). However, as the car moves backward, the system makes the character naturally transition to a backward jogging animation, adapting pose and root motion to avoid collision.}
    \label{fig:car}
\end{figure*}

Figure~\ref{fig:car} shows a car moving towards the character, while the \textbf{user simultaneously provides input to make the character walk towards the car} (e.g., pressing the \textit{down} arrow key). Initially, the character approaches the car. However, just as the character is about to collide, the animation naturally transitions to a jogging backwards motion to avoid the vehicle. Crucially, not only is the pose altered, but the character's root motion is also adapted to move backward, even though the user's target trajectory remains directed towards the car.

\subsubsection{Multi-character Interaction}
Multi-character interaction distinguishes itself from the previous experiments as each character is independently animated by our Environment-aware Motion Matching system.

For real-time interaction, each character computes obstacle penalizations by considering the expected future ellipses of other characters. Specifically, for each of its own three future ellipses (e.g., at 20, 40, and 60 frames ahead), a character only considers the \textit{corresponding} future ellipse from other agents (e.g., the 20-frame ellipse with another agent's 20-frame ellipse). This ensures consistent time horizons for the predicted interactions.

In addition, we include some subtle upper-torso motions in the animation dataset to resemble more natural interactions between agents.

\begin{figure*}
    \centering
    \includegraphics[width=1\linewidth]{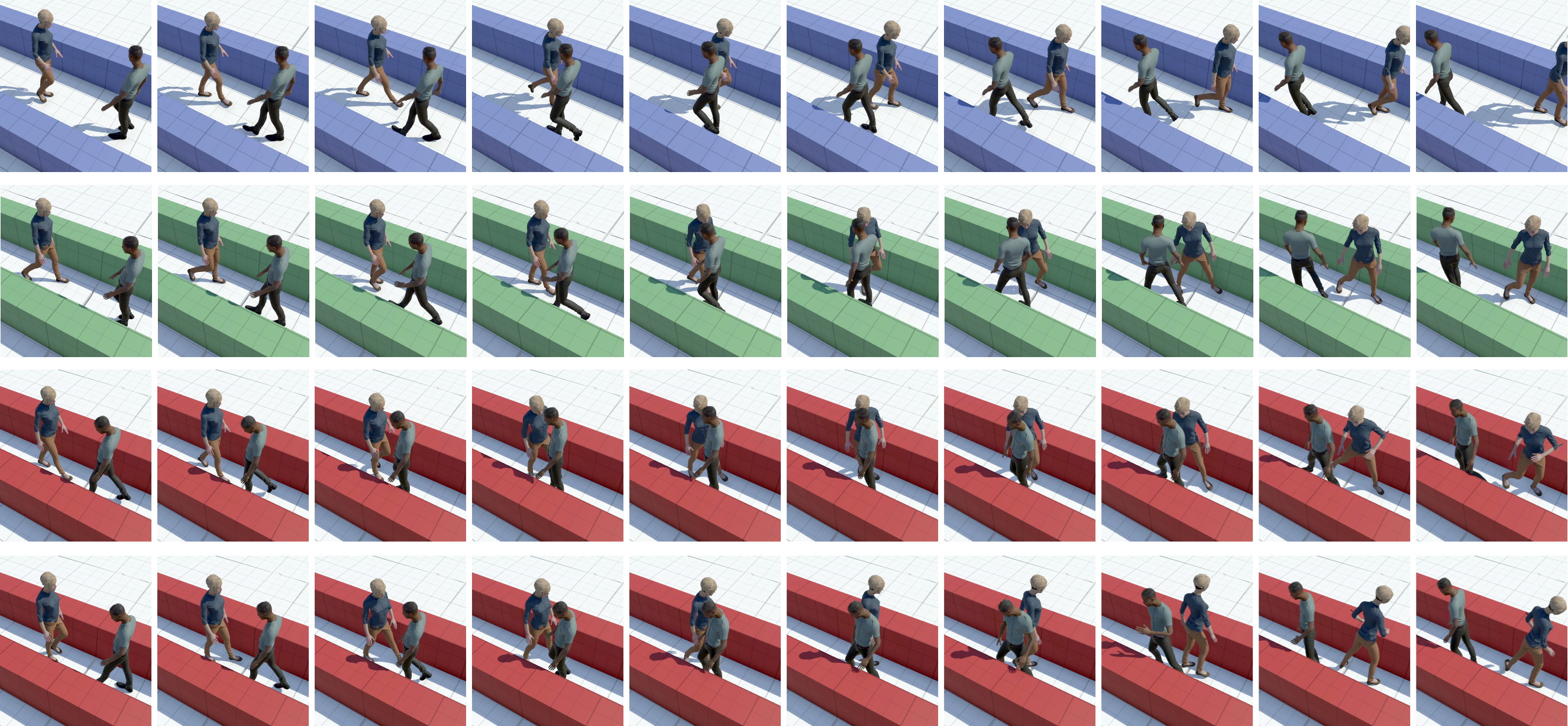}
    \caption{Multi-character interaction in corridors of varying widths. \textbf{Top (blue corridor, 1.55 m width):} Characters walk with minimal body adjustment. \textbf{Middle (green corridor, 1.20 m width):} Characters exhibit body turns to pass each other while walking. \textbf{Bottom (red corridor, 0.95 m width):} Characters carefully avoid each other. Two examples of the red corridor are provided to demonstrate the diversity of poses generated by our method.}
    \label{fig:wide_mid_narrow_corridor}
\end{figure*}

Figure~\ref{fig:wide_mid_narrow_corridor} displays two characters walking towards each other in three corridors of varying widths: (1) the blue corridor (1.55\,m width) provides ample space, allowing both characters to walk naturally with minimal torso movement; (2) the green corridor (1.20\,m width) requires both characters to turn their bodies while walking to fit past each other; (3) the red corridor (0.95\,m width) is considerably narrower, requiring characters to carefully avoid one another. We provide two examples for the red corridor to show the diversity of generated poses in such constrained situations.

\begin{figure*}
    \centering
    \includegraphics[width=1\linewidth]{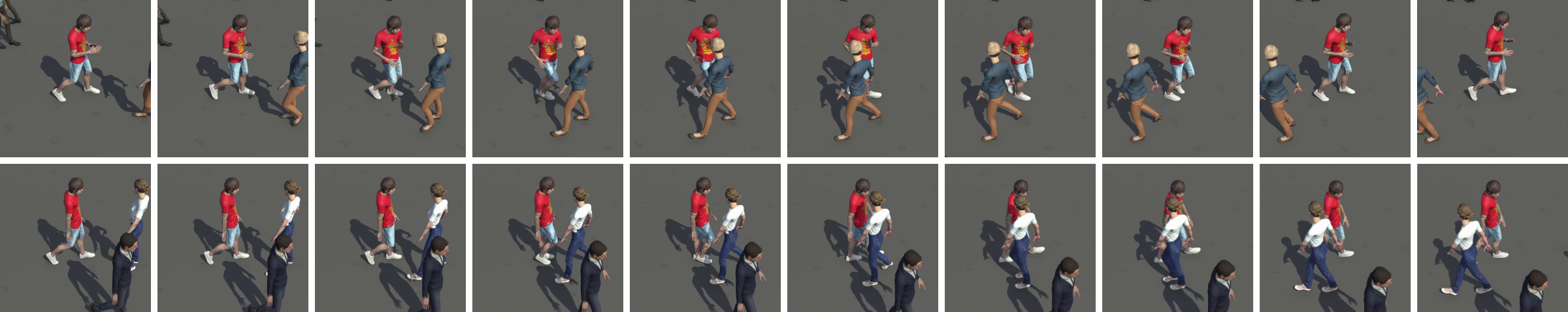}
    \caption{Interaction between agents with different speeds. \textbf{Top row:} A running character (red shirt) approaches a walking character, leading to a noticeable torso rotation for avoidance. \textbf{Bottom row:} Both characters are walking, resulting in more subtle torso rotations for collision avoidance.}
    \label{fig:agents}
\end{figure*}

Figure~\ref{fig:agents} illustrates two interaction scenarios. The first row shows a character (red t-shirt) running towards a walking character. A considerable torso rotation is observed to avoid collision, yet velocity is maintained due to sufficient available space. In the second row, both characters are walking, and the torso rotation for avoidance is more subtle, reflecting the lower speed and the less urgent need for drastic adjustments.

\subsubsection{Adding Additional Environment Features}
\label{sec:eval:height-features}
Our framework is designed with the generalization of the environment features in mind. Fundamentally, these features are used to compute penalizations that guide the search process based on the dynamically changing environment. Thus, we are not limited to encoding 2D ellipses. For instance, we extend the environment features by incorporating height information to determine whether a character should jump or crouch to avoid vertical obstacles. This involves adding two real numbers per ellipse, representing the minimum and maximum vertical components in character space. During the search, large penalties are applied when there is an overlap between the vertical segment defined by the character and that of the obstacles(computed automatically from the bounding box), effectively discarding incompatible trajectories.

\begin{figure*}
    \centering
    \includegraphics[width=1\linewidth]{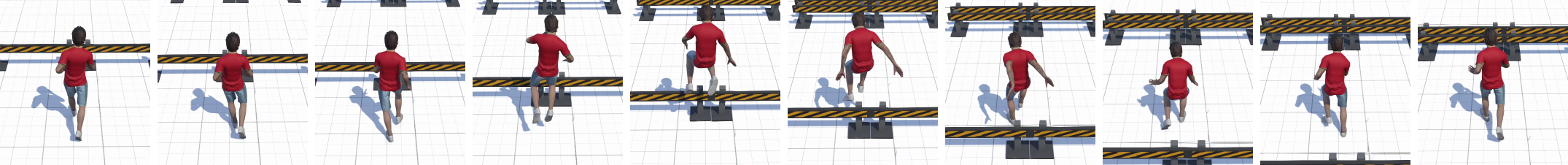}
    \caption{Character interacting with vertical obstacles using height features. The character jumps over a fence to maintain its target trajectory, demonstrating the system's ability to select appropriate vertical movements to overcome obstacles.}
    \label{fig:jumps}
\end{figure*}

\begin{figure*}
    \centering
    \includegraphics[width=1\linewidth]{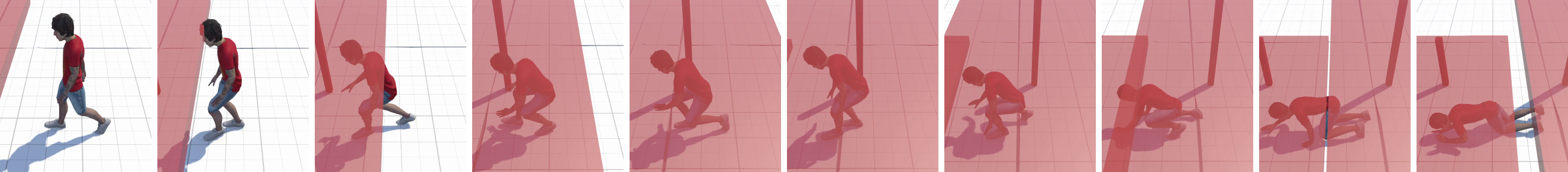}
    \caption{Character adapting to varying ceiling heights. The character progressively crouches to pass under a semitransparent red ceiling. When faced with a significantly lower ceiling, the character naturally transitions to a lying-down pose to traverse the obstacle, showcasing vertical adaptation.}
    \label{fig:crouching}
\end{figure*}

Figure~\ref{fig:jumps} illustrates a character jumping over a fence to reach its destination. This action is chosen because it better approximates the target trajectory compared to taking a detour. Figure~\ref{fig:crouching} shows a character progressively crouching to walk under a transparent red ceiling. When encountering a second ceiling at a considerably lower height, the character naturally lies down to pass underneath.

\subsubsection{Crowd Simulation Integration}
A primary objective of this work is to bridge the gap in current crowd simulation techniques, where animations are typically layered onto a predefined root motion. This common approach often leads to inconsistencies between the character's root motion and its full-body pose. Our method offers a straightforward integration into crowd simulation algorithms: these algorithms can generate global target trajectories, and our system then produces root motion that locally avoids obstacles, perfectly synchronized with realistic poses.

\begin{figure*}
    \centering
    \includegraphics[width=1\linewidth]{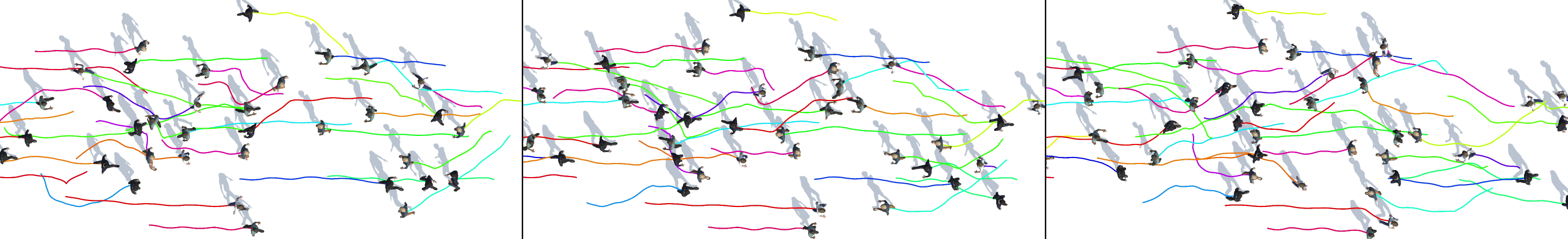}
    \caption{Our method can be trivially integrated with crowd simulation systems, allowing characters to adapt their poses and motion to the other characters. The images shows three snapshots (100 frames apart) from a simulation using a basic rule-based crowd steering. See accompanying video.}
    \label{fig:corridor-crowds}
\end{figure*}

We demonstrate this integration with a basic rule-based crowd simulation algorithm. In this setup, each agent employs a cone of vision and computes forces perpendicular to the direction towards the nearest character. This steering force is then applied to the target trajectory to avoid collisions. This basic strategy, when combined with our method, is effective for typical crowd scenarios, such as two groups of characters walking toward each other (see Figure~\ref{fig:corridor-crowds}).

\begin{figure}
    \centering
    \includegraphics[width=1\linewidth]{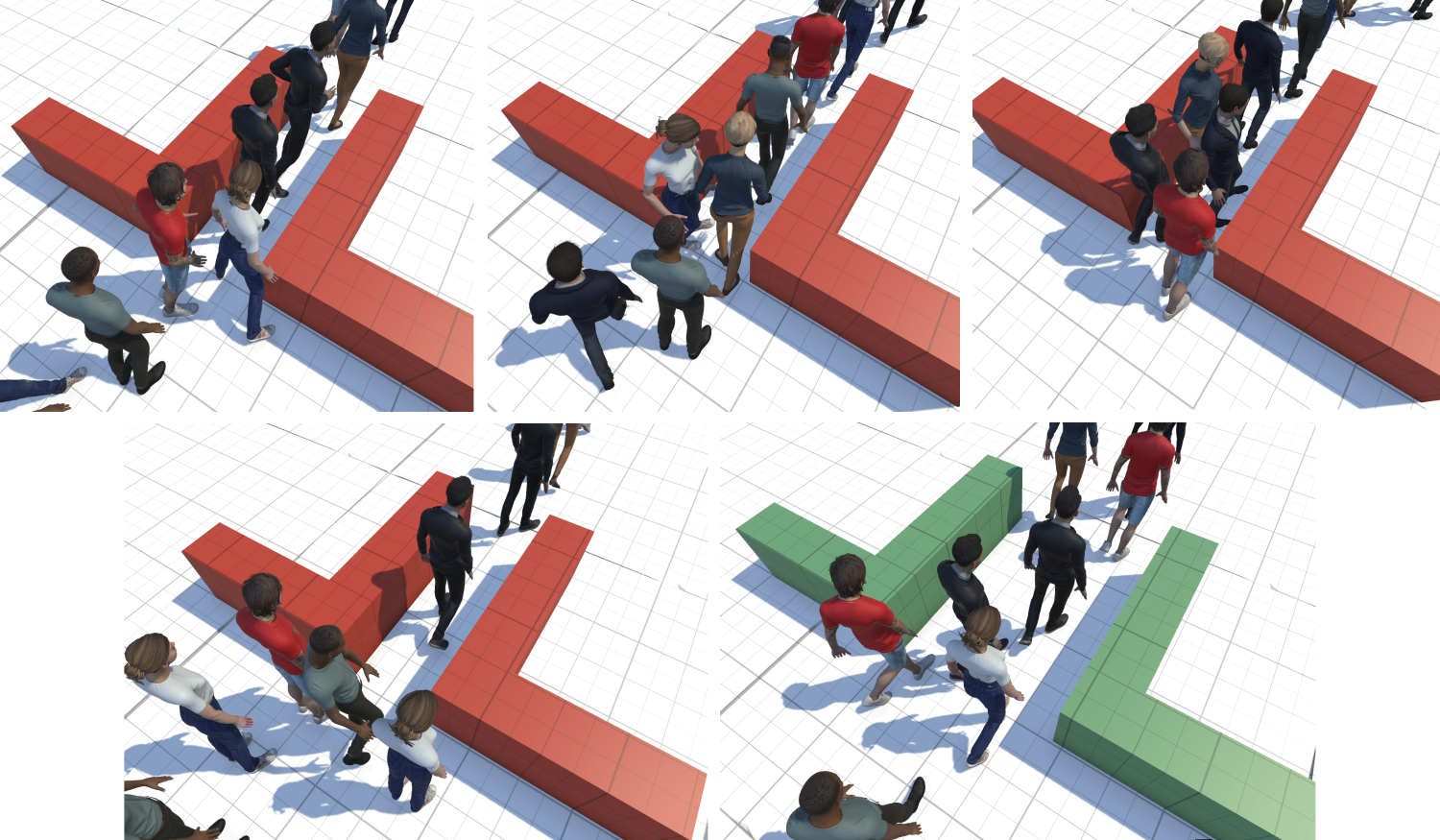}
    \caption{Crowd agents navigating doors of different widths and responsiveness settings. The green corridor (1.85\,m wide) shows characters passing with little torso rotation. The red corridor (0.9\,m wide) often requires sidestepping. \textbf{Top row:} Agents prioritize crossing quickly (high \textit{Responsiveness}), demonstrating more direct, yet still adaptive, paths. \textbf{Bottom row:} Agents are less rushed (low \textit{Responsiveness}), leading to more pronounced avoidance behaviors and larger clearance.}
    \label{fig:door-crowds}
\end{figure}

We also present another common scenario utilized in crowd simulation research. Figure~\ref{fig:door-crowds} displays two doors: a green one with an opening of 1.85\,m and a red one with an opening of 0.9\,m. Agents can be observed walking through the wider green door with minimal torso rotation. In contrast, traversing the narrower red door requires sidestepping in certain instances. The upper row of Figure~\ref{fig:door-crowds} shows examples in which agents prioritize crossing quickly (high \textit{Responsiveness}). In contrast, the bottom row illustrates characters that are not in a rush (low \textit{Responsiveness}), consequently attempting to maintain greater distance from other characters. Our method not only finds a natural coupled solution between animation and trajectory, but also produces waiting behaviors since the method provides natural idle animations when the agent encounters a bottleneck.

\subsubsection{Animation Style}
A significant advantage of our method is its rapid iteration time, allowing for quick integration of diverse animation styles and precise control over the final visual result. Changing animation styles is as simple as capturing new motion capture data and plugging it directly into the system to see it in action immediately. In this experiment, we explore three distinct styles.

\begin{figure*}
    \centering
    \includegraphics[width=1\linewidth]{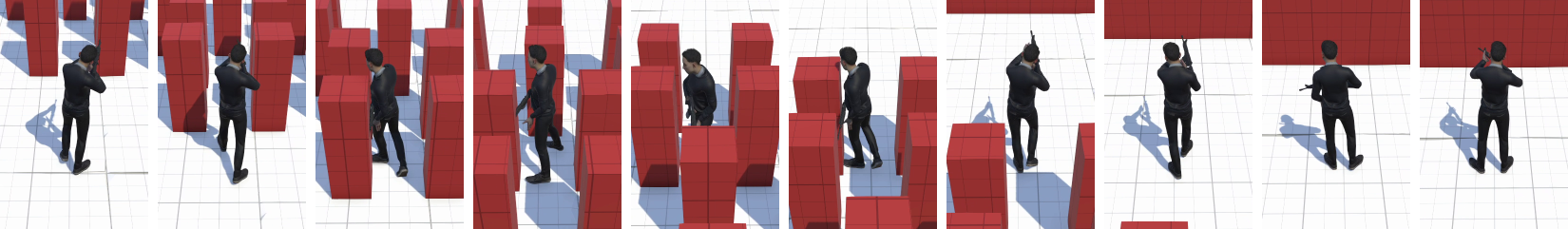}
    \caption{Character adapting pose while holding a prop weapon. The character lowers the weapon as it navigates between columns to successfully fit through the narrow gaps.}
    \label{fig:weapon-style}
\end{figure*}

\begin{figure*}
    \centering
    \includegraphics[width=1\linewidth]{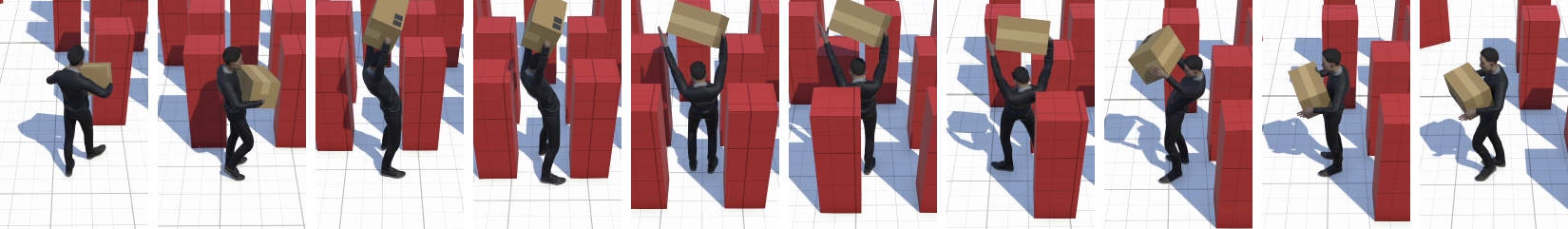}
    \caption{Character carrying a large box. The character raises the box above its head to pass between columns.}
    \label{fig:box-style}
\end{figure*}

\begin{figure*}
    \centering
    \includegraphics[width=1\linewidth]{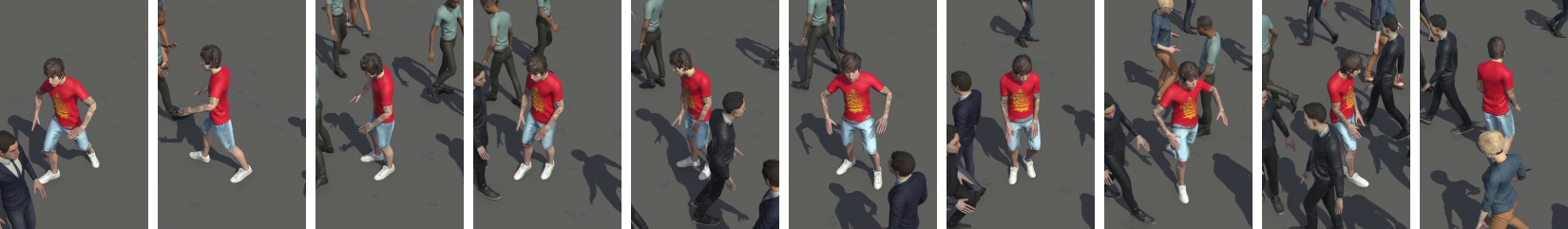}
    \caption{Example of a different locomotion style. A character typically walking with elbows raised adopts a more compact \textit{quiet} pose when near other agents to fit through spaces, demonstrating adaptation to social or spatial constraints based on animation style.}
    \label{fig:elbowsup-style}
\end{figure*}

Figure~\ref{fig:weapon-style} shows a character holding a prop weapon and adapting its body to walk between columns. The character lowers the prop when getting close to the columns to pass through. Similarly, in Figure~\ref{fig:box-style}, the character is carrying a large box and is required to raise it above its head to fit between columns. Finally, in Figure~\ref{fig:elbowsup-style}, a character that typically prefers to occupy more space by raising its elbows while walking, adopts a more \textit{quiet} pose when navigating near other agents to fit through tight spaces.

All three styles require no changes to the algorithm; only the animation data needs to be swapped to adapt the target style. For instance, in the large box example, the ellipse representing the body shape is larger when the character carries the box at its side and smaller when held overhead. This naturally forces the character to carry the box over its head when walking in tight spaces.

\subsection{Comparison vs Standard Motion Matching}
\label{sec:evaluation:comparisons}

We quantitatively (see Section~\ref{sec:evaluation:performance}) and qualitatively compare our Environment-aware Motion Matching system against standard Motion Matching to highlight the benefits of integrating environmental features. As discussed, traditional Motion Matching operates by searching an animation database for pose sequences that align with a target trajectory and maintain continuity with the current pose. However, this approach inherently disregards the character's surroundings and any multi-agent interactions. While it is theoretically possible to include additional environment-specific features (e.g., local object positions) within a Motion Matching database, such data would require laborious pre-labeling and pairing, severely limiting the database's practicality and generalizability. In contrast, our environment features do not make assumptions about the obstacles, which facilitates reusing a single sequence of poses for multiple situations.

\begin{figure}
    \centering
    \includegraphics[width=1\linewidth]{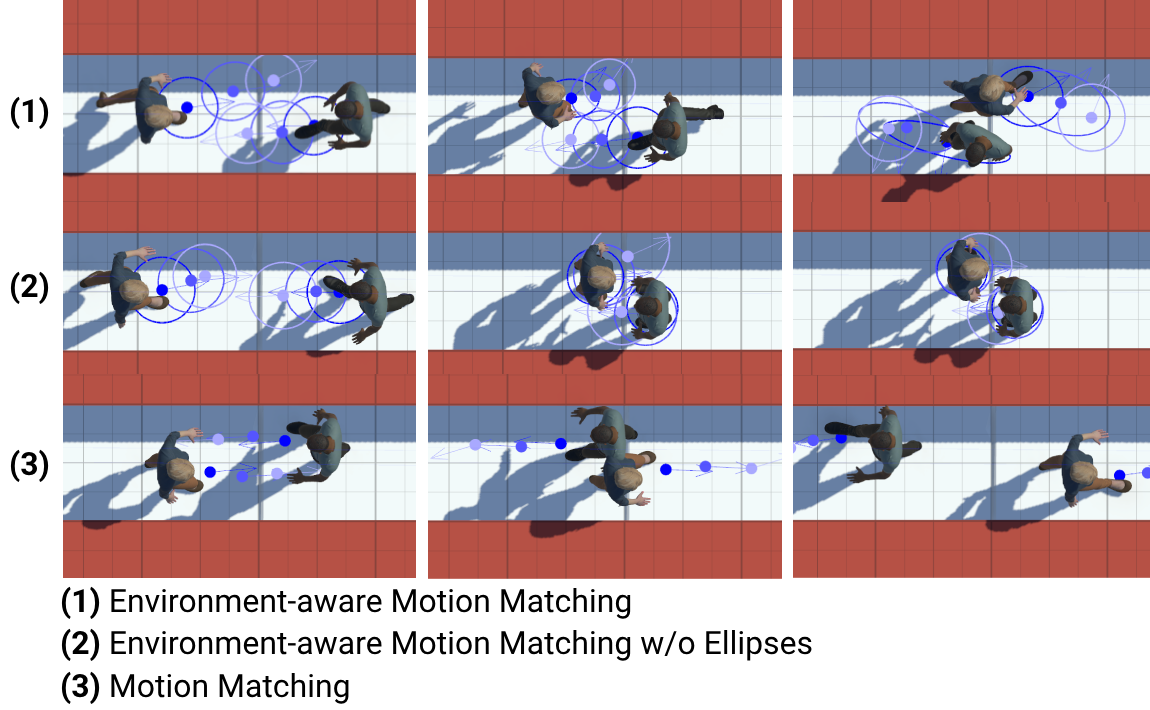}
    \caption{Comparison of body representations and environmental awareness. The top row illustrates our method using ellipses, where characters adapt their poses (e.g., sidestepping) to navigate past each other. The middle row shows our method using disks, demonstrating how characters become stuck due to the limited expressive power of disk approximation. The bottom row presents standard Motion Matching, where characters completely ignore each other and collide, highlighting its lack of environmental awareness.}
    \label{fig:emm-vs-disks-vs-mm}
\end{figure}

The impact of this fundamental difference is clearly visible in real-time scenarios. Figure~\ref{fig:emm-vs-disks-vs-mm} (bottom row) visually demonstrates how a standard Motion Matching character, lacking environmental awareness, directly collides with another agent, completely ignoring its presence.

In Figure~\ref{fig:obstacle_trajectory}, our character alters its path to navigate around a large obstacle, seamlessly integrating avoidance into its movement. Standard Motion Matching, without our environmental features, would attempt to maintain the original target trajectory, inevitably leading to a collision with such an obstacle.

\subsection{Performance and Ablation Analysis}
\label{sec:evaluation:performance}
One of the primary challenges in integrating environment features with dynamic obstacles is the linear scaling of the database traversal with the number of obstacles and poses. In Section~\ref{sec:method:optimizations}, we detailed various optimizations to achieve real-time performance, categorizing them into base optimizations and temporal coherence optimizations. We do not provide a separate evaluation for the base optimizations, as they consistently enhance performance without altering the method's core behavior. This section focuses on evaluating the temporal coherence optimizations, which often involve a trade-off between execution speed and accuracy. 

\begin{figure}
    \centering
    \includegraphics[width=1\linewidth]{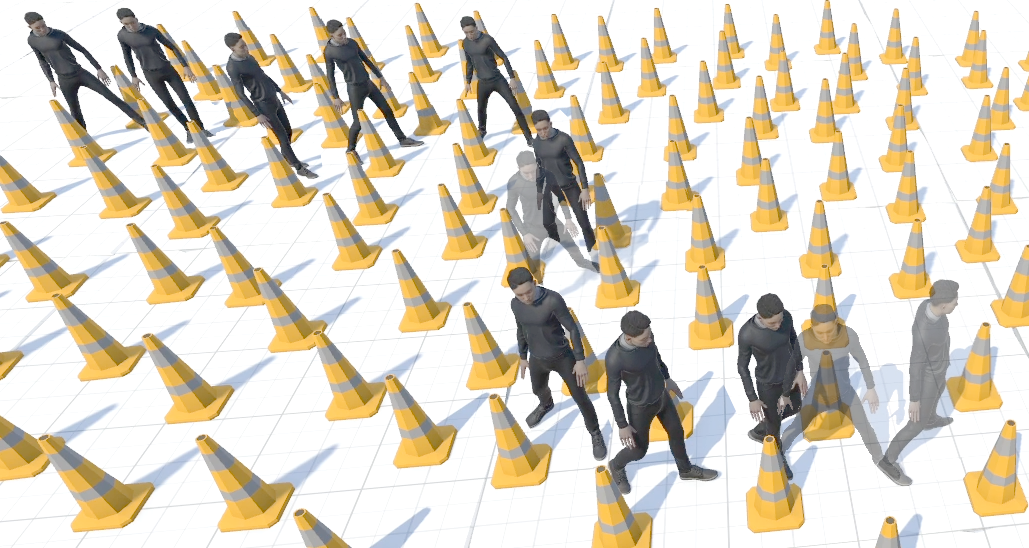}
    \caption{A part of the test environment employed for the performance and ablation analyses presented in Table~\ref{tab:performance:ablation}. Our Environment-aware system enables characters to navigate through dense arrangements of cones, dynamically adjusting their body pose to fit between tight spaces.}
    \label{fig:performance_eval}
\end{figure}

\paragraph{Experimental setup.}
We designed an experimental environment densely populated with obstacles, as depicted in Figure~\ref{fig:performance_eval}, and tasked the character controller with following a predefined path for approximately 1 minute and 30 seconds. This setup represents a stress scenario for our system, as the constant and dense obstacle environment necessitates intensive use of environment features.

\paragraph{Metrics.}
Table~\ref{tab:performance:ablation} summarizes our performance and ablation analysis. We report the average \textbf{performance} of each feature search, which is executed every 10 frames at an application frame rate of 60\,Hz. It is important to note that our search algorithm runs on the CPU within a single thread. Additionally, we report: the \textbf{diversity}, quantified as the number of distinct poses utilized; the \textbf{error}, representing the average deviation of the character position from the target path per frame in meters; and \textbf{collision time}, expressed as the average duration of each obstacle intersection.

\paragraph{Results.}
Next, we introduce and discuss each experiment. Our non-environment-aware baseline is the standard Motion Matching, which achieves the highest performance ($0.22 \pm 0.05$) due to its BVH-accelerated search. However, it does not adapt to the environment, therefore yielding a low diversity count (1997), and a significant time per collision ($0.88 \pm 0.64$). In particular, this baseline spends approximately 30\% of its time intersecting obstacles, compared to only 2\% for our complete approach. In addition to the increased collision duration, a visual inspection of the resulting animation (see accompanying videos) shows that baseline motion matching leads to arbitrary penetrations, with the character passing directly through the obstacle. In contrast, our environment-aware method significantly reduces both the duration and the extent of such penetrations.

Upon incorporating our environment-aware framework, the diversity significantly increases (3305), and the average time per collision drastically decreases ($0.05 \pm 0.04$). This is demonstrated by the non-optimized experiment, \textit{Linear Search}, which simply traverses the feature database with only base optimizations. However, this comes at a high performance cost ($9.50 \pm 4.16$).

The subsequent experiment, \textit{One Obstacle}, is designed to motivate the need to skip feature vectors during the search, rather than solely relying on further improvements to the early rejection mechanism. In this experiment, only the intersection of the first ellipse with the first obstacle is checked. This effectively represents an ideal scenario where the early rejection mechanism functions optimally for all feature vectors. Nevertheless, we observe a performance of $5.83 \pm 0.12$\,ms, which underscores the need for a mechanism to skip feature vectors.

Next, we demonstrate the importance of the temporal coherence-based optimizations. The \textit{No Adaptive} experiment removes both the minimum search stride and the adaptive threshold, with the stride adjusted solely based on the dissimilarity between the current feature vector and the best one. Conversely, the \textit{No Dissimilarity} experiment utilizes the minimum search stride and adaptive threshold while omitting the dissimilarity-based stride adjustment. Finally, the \textit{Start 0} experiment consistently begins the feature search from the first index, rather than from the previously best-matched index. In all three of these cases, performance is negatively impacted, while the trajectory error and collision metrics yield similar results.

In contrast to the \textit{No Dissimilarity} experiment, the \textit{Large v} experiment uses a large search aggressiveness parameter ($v=10$, compared to $v=5$ in other experiments), which exaggerates the effect of the dissimilarity-based stride. As a result, performance is slightly reduced, likely because the aggressive search skips good candidates early on (as indicated by the higher standard deviation). The trajectory error is also considerably impacted, as the system fails to find the most appropriate poses. This highlights the importance of balancing the v parameter to achieve both performance and accuracy.

Only when our final approach, incorporating all optimizations and denoted as \textit{Ours}, is employed, do we achieve the highest performance ($0.80 \pm 0.16$). The trajectory error is slightly higher compared to the Motion Matching baseline ($0.14 \pm 0.07$), which is an expected trade-off as the system prioritizes obstacle avoidance, sometimes at the expense of minor detours from the target trajectory.

Interestingly, our full approach exhibits the highest pose diversity (4378), even surpassing other environment-aware experiments. We attribute this to the coarse-grained search strategy. In an exhaustive traversal of the database, the system might frequently settle on sequences that offer a perfect match to the virtual environment. By introducing a coarse-grained search that occasionally skips these exact matches, the method is encouraged to explore alternative, near-optimal sequences, thereby significantly increasing the observed pose diversity.

Finally, it should be noted that our method achieves the same performance results as the base Motion Matching ($0.22 \pm 0.05$) when no obstacles are present. However, for this specific experiment, we deliberately focused on evaluating performance within a worst-case scenario.

\begin{table}
    \footnotesize
    \centering
    \caption{Performance and ablation analysis of our Environment-aware Motion Matching system. We report the average \textbf{performance} of the feature search (in milliseconds, executed every 10 frames), \textbf{diversity} (number of distinct poses used), \textbf{trajectory error} (average deviation from target path in meters), and \textbf{collision time} (average time of obstacle intersection in seconds). Arrows indicate whether a lower ($\downarrow$) or higher ($\uparrow$) value is better. Results demonstrate that our approach achieves near real-time performance with high pose diversity and minimal collisions compared to the Motion Matching baseline and various optimization configurations, especially in a worst-case dense obstacle environment (Figure~\ref{fig:performance_eval}).}
    \begin{tabular}{lcccc}
        \toprule
            \multirow{2}{*}{Ablation} & Performance $\downarrow$ & Diversity $\uparrow$ & Error $\downarrow$ & Collision time $\downarrow$ \\
            & (ms) & (count) & (m) & (s) \\
                \midrule
            \textit{Motion Matching}& \barR{0.22} $\pm 0.05$ & \barD{1997} & \barE{0.09} $\pm 0.04$ & \barC{0.88} $\pm 0.64$ \\
            \textit{Linear Search}  & \barR{9.50} $\pm 4.16$ & \barD{3305} & \barE{0.12} $\pm 0.06$ & \barC{0.05} $\pm 0.04$ \\
            \textit{One Obstacle}   & \barR{5.83} $\pm 0.12$ & \barD{2103} & \barE{0.09} $\pm 0.04$ & \barC{0.79} $\pm 0.67$ \\
            \textit{Disk}           & \barR{0.90} $\pm 0.22$ & \barD{3838} & \barE{0.29} $\pm 0.17$ & \barC{0.37} $\pm 0.29$ \\
            \textit{No Adaptive}    & \barR{2.40} $\pm 0.70$ & \barD{3729} & \barE{0.13} $\pm 0.07$ & \barC{0.08} $\pm 0.08$ \\
            \textit{No Dissimilarity}& \barR{1.62} $\pm 0.17$ & \barD{3608} & \barE{0.13} $\pm 0.08$ & \barC{0.12} $\pm 0.10$ \\
            \textit{Start 0}        & \barR{1.02} $\pm 0.20$ & \barD{3807} & \barE{0.16} $\pm 0.10$ & \barC{0.06} $\pm 0.12$ \\
            \textit{Large} $v$ & \barR{0.88} $\pm 0.35$ & \barD{4389} & \barE{0.37} $\pm 0.34$ & \barC{0.12} $\pm 0.10$ \\
            \textit{Ours}           & \barR{0.80} $\pm 0.16$ & \barD{4378} & \barE{0.14} $\pm 0.07$ & \barC{0.08} $\pm 0.05$ \\
         \bottomrule
    \end{tabular}
    \label{tab:performance:ablation}
\end{table}

\subsection{Animation Database Scalability Analysis}
\label{sec:evaluation:scalability}

\begin{figure}
    \centering
    \includegraphics[width=1\linewidth]{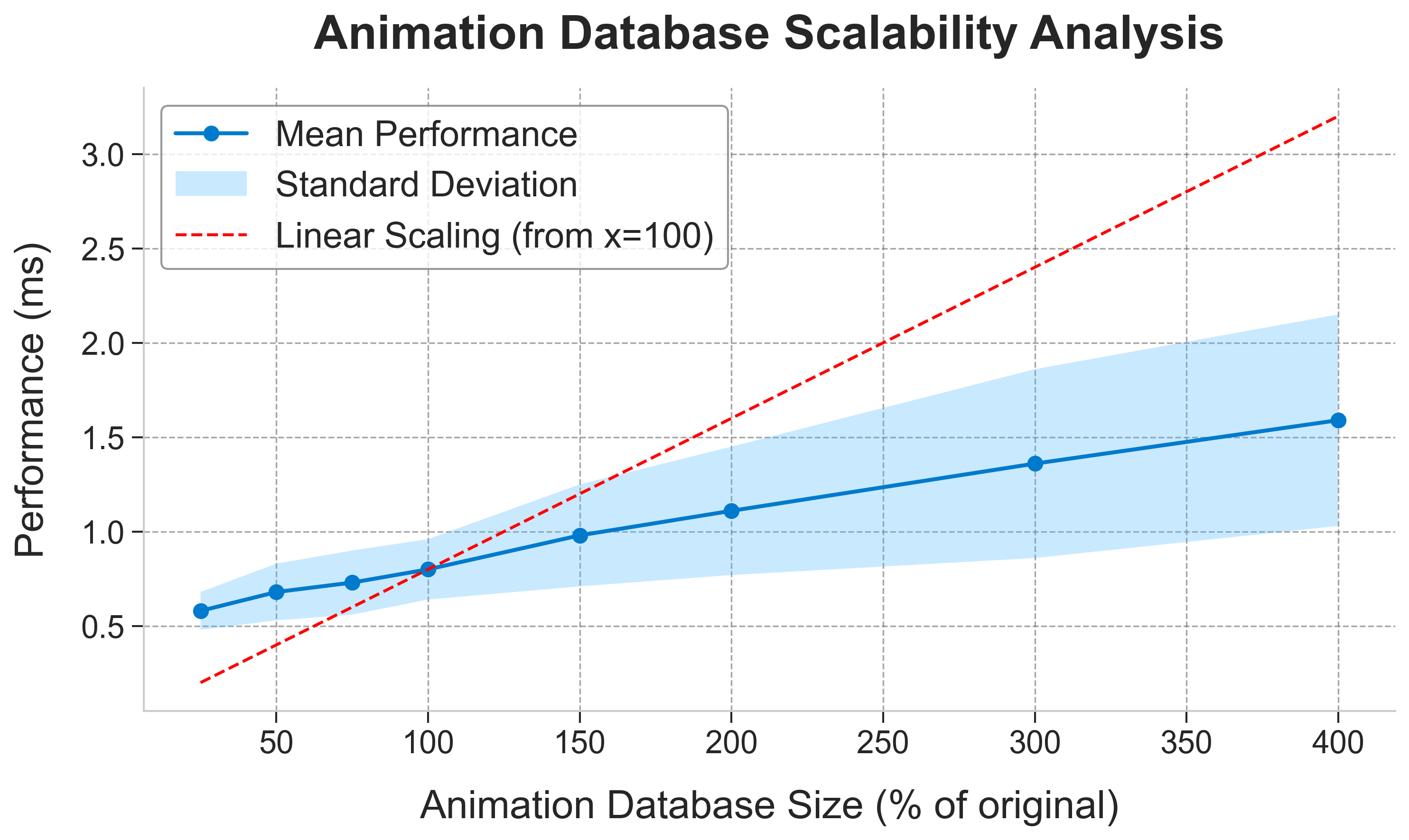}
    \caption{Animation Database Scalability Analysis. The graph displays the mean performance (in milliseconds) and standard deviation as a function of the animation database size, where the original database used in Table~\ref{tab:performance:ablation} represents 100\%. The expected linear scaling reference (from x=100\%) is included to illustrate how the increase in performance cost with database size has a significantly reduced linear slope.}
    \label{fig:scalability_eval}
\end{figure}

In this section, we analyze the impact of the animation database size on the performance, following the experimental setup detailed previously. Figure~\ref{fig:scalability_eval} presents the mean performance and standard deviation for various dataset sizes. We use the database in Section~\ref{sec:evaluation:performance} as our 100\,\% reference point (containing 101,026 poses). We then generated datasets of different relative sizes by trimming or expanding this original database: 25\,\% (26,455 poses), 50\,\% (48,575 poses), 75\,\% (75,341 poses), 150\,\% (154,214 poses), 200\,\% (209,152 poses), 300\,\% (301,794 poses), and 400\,\% (411,280 poses).

As shown in Figure~\ref{fig:scalability_eval}, our method demonstrates considerable scalability with increasing animation database size. The red dashed line indicates the expected linear growth, referenced from the 100\,\% database size. In other words, if we duplicate the number of poses in the database, we would expect the feature search to take twice the time. Our method consistently shows a performance increase with a significantly reduced linear slope as the dataset grows. Conversely, when the database size is reduced, performance remains relatively constant. This latter effect is primarily due to constant costs associated with base optimizations, such as the initial search for nearby obstacles, which constitute a significant portion of the performance overhead when the dataset is small.

We attribute the reduced linear slope growth to two primary factors: First, the early rejection mechanism effectively discards a large portion of the database once a sufficiently good candidate feature vector is identified. Second, while the animation database size can grow linearly, the number of relevant obstacles for an agent remains bounded by its immediate surroundings. Nonetheless, the standard deviation of performance increases proportionally with the animation database size, which could complicate maintaining a constant frame rate. This challenge could potentially be addressed by implementing a budget-based search, where the search is terminated once a predefined maximum frame time is reached.

Finally, regarding multi-agent environments, we do not provide a specific scalability analysis since the reported performance scales linearly with the number of agents. However, note that each feature search is performed every 10 frames (for a 60 Hz application). This allows for distributing the computational load of searches across multiple frames, thereby accommodating a larger number of agents.

\subsection{Disk-based vs Ellipse-based Body Representation}
\label{sec:evaluation:ellipse}

A critical aspect of our real-time environment-aware character controller is the selection of environment features used to approximate the character's shape. In this work, we chose 2D ellipses due to their compact representation (requiring only 3 real numbers per ellipse) and their ability to differentiate various motions, such as torso rotation and sidestepping. To evaluate the necessity of ellipses, we compare our framework using ellipses against a configuration encoding disks as environment features. 

A main limitation of disks is their inability to distinguish between different types of motion. Figure~\ref{fig:emm-vs-disks-vs-mm} illustrates the difference between our method using ellipses (first row) and using disks (second row). When two characters walk towards each other in a narrow passage, our ellipse-based method naturally finds a sequence of poses (e.g., sidestepping) that allows them to fit within the corridor and avoid collision. This is not possible with disks, causing the characters to get stuck, as no motion other than an idle pose can proceed without collisions.

We also quantitatively evaluate these observations in Table~\ref{tab:performance:ablation}, specifically in the row labeled \textit{Disk}. When using disks, we observe the highest trajectory error. This occurs because the character is frequently unable to navigate between obstacles and remains stuck until the target destination is sufficiently far to overcome the obstacle penalizations. When such movement does occur, the character inevitably intersects with obstacles, as evidenced by the large average time per collision.
\section{Limitations and Future Work}
\label{sec:limitations}

In this work, we have presented a framework for enabling environment-aware motion matching by incorporating a specialized type of feature that seamlessly integrates with existing query features, allowing for greater flexibility in analyzing the character's surroundings. While we demonstrate its efficacy by approximating the character's projected body shape using 2D ellipses, certain aspects could be further refined. Our current representation only considers the skeleton and does not account for the character's actual mesh or skin. For applications demanding higher geometric accuracy, this could be easily addressed during the preprocessing stage by projecting all vertices of the character mesh onto the feature representation. Furthermore, other body shape representations, such as convex hulls or even full 3D representations, could be employed to better fit complex character geometries. Similarly, our method cannot strictly guarantee collision-free motion, as the poses for trajectory and environment features are sampled at discrete intervals (333\,ms apart), which might allow for fleeting penetrations between samples.

Another important consideration when integrating environment-aware techniques is the responsiveness to user input. Applications requiring very high control over character movement, such as fast-paced video games, often prioritize direct control over animation root motion to enable fine-grained character manipulation. While the \textit{Responsiveness} control allows for adjusting this trade-off, our method will inherently generate poses that may deviate from the precise target trajectories to accommodate environmental constraints. A comprehensive and diverse animation database is crucial for minimizing these deviations and ensuring natural-looking adaptations.

\section{Conclusions}
\label{sec:conclusions}

In this paper, we have presented a data-driven method that enables the animation of characters that react naturally to environmental conditions, including dynamic obstacles and other characters.
We have introduced a framework that clearly separates query features, which are matched against the user input, from environment features, which facilitate collision-free locomotion. The required input (essentially the intended short-term trajectory) is flexible enough to be derived from a variety of modalities, ranging from simple keyboard control to a crowd steering engine.

A major benefit of our approach is that it captures the natural relationship between human poses and trajectory, a relationship largely absent in crowd simulations and often responsible for visibly inconsistent motion. This capability allows our animated characters to respond naturally to a dynamic environment, transitioning between different animations. These responses range from subtle pose adjustments, such as torso rotation, to more significant locomotion changes, such as sidestepping, crouching, or jumping. In addition to automatically selecting the appropriate locomotion modality, our method also adjusts the trajectory based on environmental constraints. This significantly simplifies user input, as our search strategy handles the selection of the most appropriate motion. This includes generating detours to avoid collisions and even transitioning from an idle state to an escape locomotion when an obstacle approaches the character, also in the absence of user input.

We have shown that a naive implementation of an environment-aware motion search introduces substantial computational overhead. To address this, we have developed a comprehensive set of optimizations that achieve a tenfold speed-up over a linear search. Furthermore, the search aggressiveness parameter $v$ enables the fine-tuning of the quality-performance trade-off per character, paving the way for level-of-detail strategies, such as using faster searches for distant characters. This adaptability makes our approach particularly suitable for crowd simulation. In addition, the memory requirements of our approach are highly competitive, requiring approximately 50 MB (uncompressed) for our motion capture database, including both the pose data and the extracted features. 

Unlike reinforcement learning approaches, our method guarantees natural motion regardless of user input or environmental conditions. Moreover, since our approach requires no training, adding support for new locomotion modalities (such as jogging, sprinting, hopping, crouching, crawling, or sliding) and styles (such as injured, exhausted, staggering, or sneaking) is as simple as extending or changing the animation database. Because our method does not require labeling of mocap data, it enables rapid iteration cycles for fine-tuning the animations to meet specific needs.

\section*{Code and data}
The source code, animation databases (184,553 poses $\sim$ 50\,min), and supplementary material used in this paper can be found at:
\newline
\href{https://upc-virvig.github.io/Environment-aware-Motion-Matching}{https://upc-virvig.github.io/Environment-aware-Motion-Matching}

\begin{acks}
This work has received funding from MCIN/AEI/10.13039/50110001 1033/FEDER, UE (Spain) in the framework of the project PID2021-122136OB-C21, and with the support of the Department of Research and Universities of the Government of Catalonia (2021 SGR 01035).
Jose Luis Ponton was funded by the Spanish Ministry of Universities (FPU21/01927).
We thank the 2025 Bellairs Workshop on Computer Animation, where the initial ideas for this research took shape.
\end{acks}

\bibliographystyle{ACM-Reference-Format}
\bibliography{References}

\end{document}